\documentclass[%
twocolumn,10pt,
superscriptaddress,
nofootinbib,
 amsmath,amssymb,
 aps,prc,
]{revtex4-2}

\usepackage{graphicx}% Include figure files
\usepackage{dcolumn}% Align table columns on decimal point
\usepackage{bm}% bold math
\usepackage{cancel}
\usepackage{mathtools}
\usepackage{dsfont}
\usepackage{float}
\usepackage[section]{placeins} %Ensure figures appear in the correct sections
\usepackage[utf8]{inputenc}
\usepackage[dvipsnames]{xcolor}
\usepackage{subfigure}
\usepackage{hyperref}
\hypersetup{
    colorlinks=true,
    citecolor=OliveGreen,
    urlcolor=cyan
}

\usepackage{comment}

\begin{document}

\preprint{APS/123-QED}

\title{The Non-Abelian Casimir Effect for Plates, Symmetrical Tube and Box on the Lattice}

\author{B.\ A.\ Ngwenya}
\email{ngwble001@myuct.ac.za}
\affiliation{%
 Department of Physics, University of Cape Town\\
 Private Bag X3, Rondebosch 7701, South Africa
}%
\author{A.\ Rothkopf}%
 \email{akrothkopf@korea.ac.kr}
\affiliation{%
 Faculty of Science and Technology, University of Stavanger\\
 4021 Stavanger, Norway
}%
\affiliation{Department of Physics, Korea University, Seoul 02841, Republic of Korea}
\author{W.\ A.\ Horowitz}%
 \email{wa.horowitz@uct.ac.za}
\affiliation{%
 Department of Physics, University of Cape Town\\
 Private Bag X3, Rondebosch 7701, South Africa
}%
\affiliation{
Department of Physics, New Mexico State University, Las Cruces, New Mexico, 88003, USA
}

\date{\today}

\begin{abstract}
We present non-perturbative results of the Casimir potential in non-abelian SU(3) gauge theory in (2+1)D and (3+1)D  in the confined and deconfined phase. For the first time, geometries beyond parallel plates in (3+1)D are explored and we show that the Casimir effect for the symmetrical tube and symmetrical box is attractive. The Casimir potential for the tube differs from the massless non-interacting scalar field theory prediction, where a repulsive Casimir potential is expected. Unlike the parallel plate geometry where the plate-size is fixed, in the case of the tube and box, the sizes of the faces forming the walls of the geometries changes with separation distance. We propose various methods that can be used to account for the energy contributions from creating the boundaries. We show that increasing the temperature from a confined to a deconfined phase does not alter the Casimir potential. This observation is consistent with prior work suggesting that the region inside the walls is a boundary induced deconfined phase.
\end{abstract}
\maketitle

\section{\label{Intro}Introduction}
In quantum field theory, fluctuating quantum fields in a vacuum are represented by a divergent sum of normal modes, which appears as an infinite amount of energy in free space, the zero-point energy. This energy can be measured by imposing different boundary conditions on the fields, which modify the vacuum properties and thus reveal zero-point energy differences. 

These energy differences were first described by Casimir \cite{Casimir:1948dh, Casimir:1947kzi}. Lamoreaux \cite{Lamoreaux:1997} provided first experimental evidence for these zero-point fluctuations. In its initial formulation, the effect was presented for a parallel plate configuration in quantum electrodynamics (QED), stating that if two identical neutral ideally conducting plates are placed at some distance $d$ apart in a vacuum, the plates would experience a small (on the order of $\sim 10^{-3}$ Nm$^{-2}$ for plates placed $\sim 1$ $\mu$m apart), but finite attractive pressure towards each other. 

Qualitatively, this can be understood as follows; the presence of the neutral conducting plates in the vacuum imposes boundary conditions on the stationary modes of the electromagnetic field in-vacuum. However, between the plates, the long wavelengths ($\lambda >d$) modes get frozen out, thus limiting the number of modes that can exist within the cavity between the plates as opposed to the number of modes in free space (outside the plates). The limited number of modes between the plates correspond to a lower vacuum energy density in that region, leading to a pressure difference and subsequently the attractive force. %When the separation distance between the plates is increased, the normal modes of `longer' wavelengths are no longer frozen out. The energy density in the region between the plates increases and the magnitude of the attractive force is reduced.

In abelian gauge theories, studies of the Casimir effect have been extended to various geometries and generalised to materials with arbitrary dielectrics and rough surfaces at finite temperature \cite{Lifshitz:1956zz}. However, such geometries cannot be easily tested experimentally nor computed analytically, with recent works exploring machine learning techniques \cite{Chernodub:2019kon}. %An example is the geometry of a large plate and a conducting spherical shell of radius $r >> d$ \cite{Bordag:2001qi, Chan:2001zzb}. %This geometry has experimental motivations introduced by the perfect alignment difficulties when placing two large plates at distances on the order of microns apart and is studied through employing the proximity force approximation \cite{Blocki:1977zz, Bimonte:2017ahs}. %Despite having a framework to perform Casimir studies with real materials, some difficulties still arise and need to be accounted for.  At finite temperatures, the fields will exhibit thermal fluctuations around the field expectation values in thermal equilibrium. 
Thermal fluctuations result in radiation pressure which adds thermal corrections to the Casimir force \cite{Mitter:1999hu, Decca:2003td,Ghisoiu:2010fga}. The thermal wavelength is given by, $\lambda_T = \hbar c/(2\pi k_BT)$ and $\lambda_T \simeq 1.2$ $\mu$m at room temperature. Hence the relevant scale where thermal fluctuations become important is at separation distances, $d \gtrsim 1$ $\mu$m  where the wavelength of such fluctuations can fit inside the contained geometry. 

In the above-mentioned generalisation, Lifshitz also predicted that there is a geometrical element to the Casimir effect, where some configurations result in a repulsive force. An example of such a configuration with a repulsive potential is the double spherical cavity where an inner conducting spherical shell of radius $r$ is surrounded by an outer spherical shell with radius, $R\to \infty$ where the outer boundary is removed to spatial infinity \cite{Boyer:1968uf,Balian:1977qr, Milton:1999ge}. A repulsive force has also been observed for a symmetrical tube in a weakly coupled, massless non-interacting scalar field computed in Ref.\ \cite{Mogliacci:2018oea} where finite volume effects on the equation of state are studied using Dirichlet boundary conditions. This study suggests that the resulting zero-temperature Casimir pressure is \textit{attractive} for a parallel plate and box configurations, and \textit{repulsive} for a tube configuration in the scalar theory. Understanding whether such a geometric dependence of the Casimir force exists in non-abelian gauge theories motivates this present work.

Other studies in QED have also revealed interesting effects resulting from the presence of boundaries, such as the Scharnhorst effect \cite{Scharnhorst:1990sr,deClark:2016mvw}, where radiative corrections for parallel plate geometry lead to photons between the plates (moving perpendicular to the plates) travelling faster than the vacuum speed of light, $c$. %Such a theoretical prediction could suggest that imposing the Casimir boundaries on the QED vacuum amplifies some frequencies in the medium cavity between the plates. This effect remains unobserved experimentally, hence the result could be plagued by theoretical errors which introduce the need to explore non-perturbative methods to compute these radiative corrections in field theories.
The abelian Casimir effect has also been studied on the lattice in an attempt to overcome inaccuracies due to e.g., thermal corrections in perturbative analytic methods for general geometries which are usually formulated using simplifications such as fixed boundary conditions. See, for example Ref.\ \cite{Bordag:2001qi} where Green's functions are used to describe the boundary effects on vacuum fluctuations in quantum field theory and Ref.\ \cite{Markov:2006js, Pavlovsky:2009kg} where the Chern-Simons action is employed.

These lattice simulations have been extended to \textit{compact} QED which exhibits properties, such as confinement in (2+1)D that make it a valuable starting point before addressing non-abelian gauge theories. We refer the reader to Ref.\ \cite{Chernodub:2016owp, Chernodub:2017mhi} for the case of parallel conducting wires in (2+1)D and Ref.\ \cite{Chernodub:2022izt} for parallel conducting plates in (3+1)D compact U(1) gauge theory using ideally conducting parallel metallic boundaries. In these studies, the resulting action describes photon dynamics which characterise the perturbative effects and the formation of abelian monopole dynamics describing the non-perturbative effects. The monopoles lead to mass-gap generation which screens the Casimir potential at large separation distances. At short separation distances, the monopole density between the boundaries is diminished into a dilute gas of monopole-antimonopole pairs and the region between the wires/plates goes through an induced deconfining phase transition.

\begin{figure}[!htbp]
    \centering
    \subfigure[Parallel Plates\label{subfig:plates_geometry}]{{\includegraphics[width=0.9\linewidth]{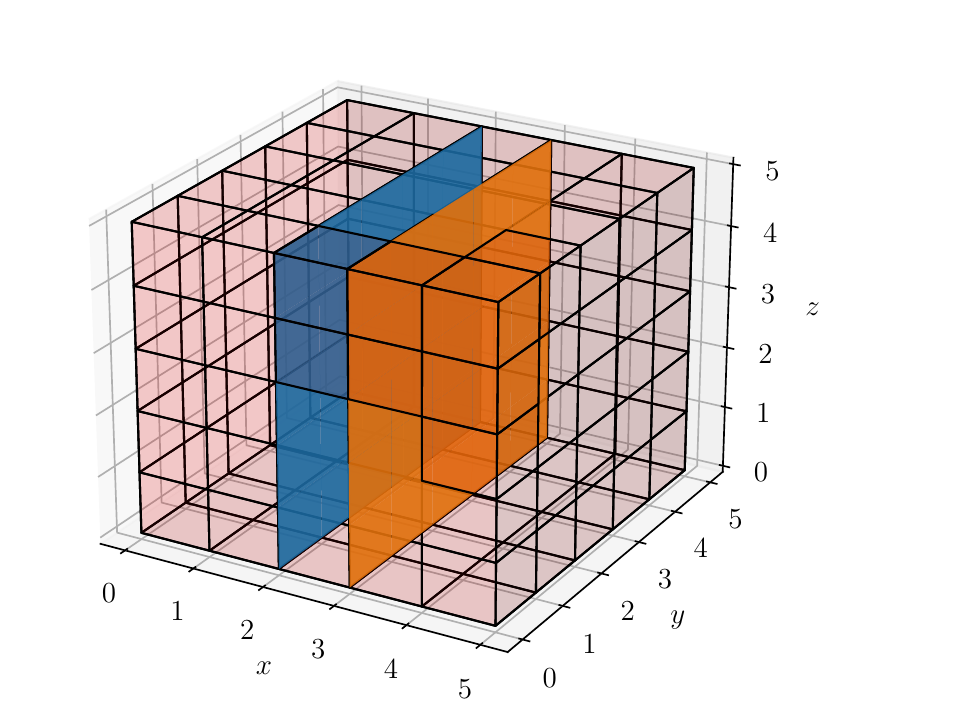} }}%
    \hspace{1em}
    \subfigure[Symmetrical Tube\label{subfig:tube_geometry}]{{\includegraphics[width=0.9\linewidth]{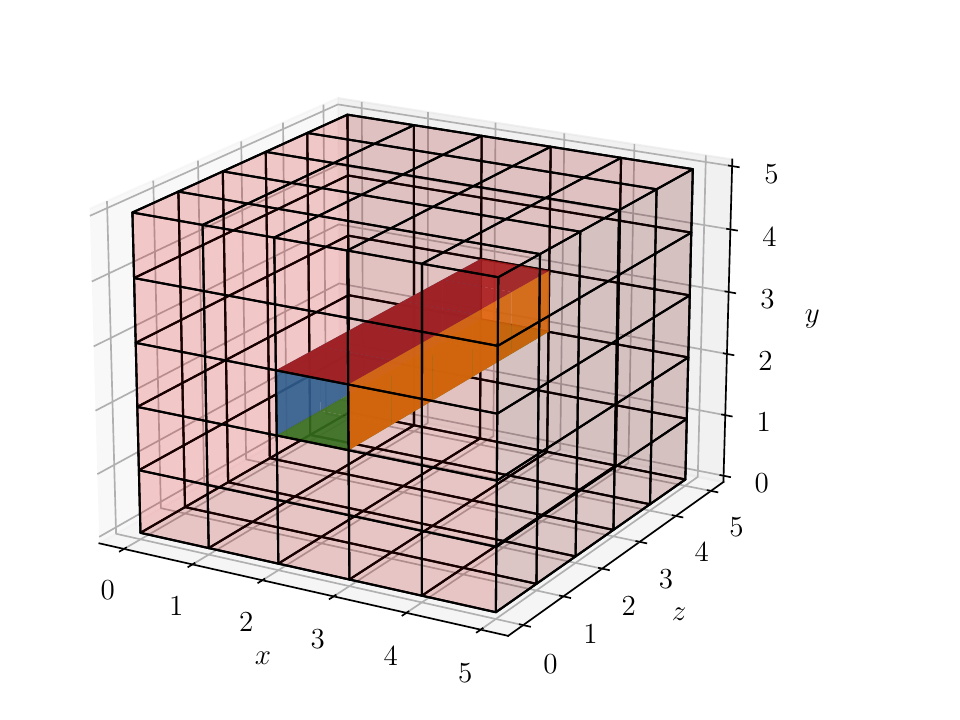} }}
    \hspace{1em}
    \subfigure[Symmetrical Box\label{subfig:box_geometry}]{{\includegraphics[width=0.9\linewidth]{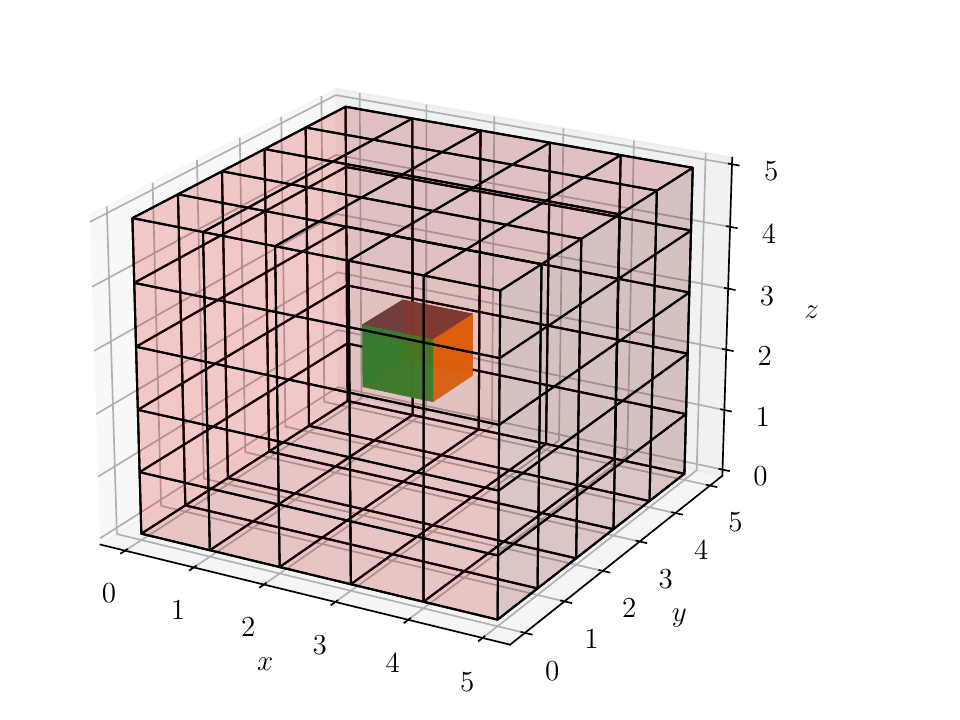} }}
    %\hspace{1em}
    \caption{Geometry of chromoelectric boundary conditions on the geometries of parallel plates, symmetrical and box in a $(3+1)$D cubic lattice.}%
    \label{fig:geometries}%
\end{figure}

%Extensive studies of the Casimir effect have been performed with abelian gauge fields because the weakly coupled theory is well-understood and the boundary conditions associated with electromagnetic fields can be experimentally controlled with great precision. In addition, radiative corrections to the Casimir energy due to interactions of virtual photons with virtual fermions can be calculated using perturbative techniques. The Casimir effect has also been studied in non-abelian gauge theories, which are relevant in the description of the strong and weak interactions.\\ 

Non-abelian Yang-Mills theories are more complicated than their abelian counterparts since they are non-linear, i.e.\ self interacting. Let us briefly review the formulation and findings of Ref.\ \cite{Chernodub:2018pmt}, where the Casimir effect in non-abelian gauge theory is studied on the lattice for perfectly conducting static parallel wires in (2+1)D SU(2) at zero temperature using chromoelectric boundaries\footnote{On the lattice, this requires the coupling at the surface of the perfect conductor material to be different from the vacuum inverse coupling, which corresponds to increasing the lattice coupling at the boundaries by a Lagrange multiplier $\lambda$, such that the tangential field component vanishes, i.e.\ Tr$[P] \to \mathds{1}$.}. The resulting attractive Casimir potential was found to be described by the empirical fitting function,
\begin{equation}
    %V_{\text{Cas}}^{\text{fit}} (R) = -(N_c^2-1)\frac{\zeta(3)}{16\pi} \frac{1}{R^{(\nu+2)}\sigma^{(\nu+2)/2}} e^{-M_{\text{Cas}}R},
    V_{\text{Cas}}^{\text{fit}} (R) = -(N_c^2-1)\frac{\zeta(3)}{16\pi} \frac{1}{R^{(\nu+2)}\sigma^{(\nu+1)}} e^{-M_{\text{Cas}}R},
    %V_{\text{Cas}}^{\text{fit}} (R) = -(N_c^2-1)\frac{\zeta(3)}{16\pi} \frac{1}{R^{\nu}\sigma^{(\nu-1)}} e^{-M_{\text{Cas}}R},
    \label{eqn:vcas_fit}
\end{equation}
with fit parameters, $\nu$ and $M_{\text{Cas}}$, where $\sigma$ is the string tension and $N_c$ is the number of colours. The parameter $\nu$ describes the anomalous scaling dimension of the potential at short separation distances compared to the expected tree-level behaviour,
\begin{equation}
    V_{\text{Cas}}^{\text{tree}} (R) = -(N_c^2-1)\frac{\zeta(3)}{16\pi} \frac{1}{R^2},
    \label{eqn:vcas_tree}
\end{equation}
describing the Casimir energy of $(N_c^2-1)$ non-interacting copies of a monopole-free $U(1)$ gauge theory analogous to a free scalar field with Dirichlet boundary conditions \cite{Ambjorn:1981xw}. The authors of \cite{Chernodub:2018pmt} concluded that similarly to the finite volume geometry and the formulation in compact QED, this Casimir geometry is bound to induce a smooth deconfining phase transition between the wires. We explore the presence of this smooth deconfining phase transition for other geometries.

The zero temperature Yang-Mills theory exhibits non-perturbative dynamic mass-gap generation in both (2+1)D and (3+1)D, similarly to compact QED. This non-perturbative mass-gap generation results in an effective screening of the Casimir potential at large separation distances quantified by the parameter $M_{\text{Cas}}$ (the Casimir mass) in the exponential of Eq.\ (\ref{eqn:vcas_fit}). The mass-dependent exponential decay of the Casimir potential is consistent with the Casimir energy decay for a massive scalar particle obtained in a gauge-invariant Hamiltonian formulation of (2+1)D non-abelian gauge theories \cite{Karabali:2018ael}. 

Interestingly, the resulting Casimir mass corresponding to the zero-energy of the (2+1)D gauge theory with perfect conductor parallel wires \cite{Chernodub:2018pmt},
\begin{equation}
    M_{\text{Cas}}^{\infty} = 1.38(3)\sqrt{\sigma} < M_{0^{++}} = 4.718(43)\sqrt{\sigma}
\end{equation}
is less than the lowest $0^{++}$ glueball mass in (2+1)D SU(2) gauge theory \cite{Teper:1998te}. In principle, the lowest glueball mass, $M_{0^{++}}$ should be the lowest mass in the system, hence this result suggests that the dominant degrees of freedom in the non-abelian theory are a lighter gluonic state, possibly forming a `light-gluon' plasma due to the boundary-induced deconfining phase transition in the region between the wires.

This low Casimir mass has also been obtained in an alternative study employing analytical methods to show that the propagator (viewed in terms of traversing from one boundary to the other and back) of the gauge-invariant gluon field in SU(2) corresponds to the propagator of a massive scalar field in (2+1)D, with a mass corresponding to the Casimir mass \cite{Karabali:2018ael}. The same study also shows that the (2+1)D Casimir mass has physical implications for the (3+1)D theory. It is shown that the Casimir mass in two-dimensions is equal to the high temperature magnetic mass due to screening effects in the three-dimensional theory. Hence studies of the Casimir effect in (2+1)D gauge theories can be used as a probe of the magnetic screening mass of a pure gauge QCD plasma. 

In motivation of our study of the Casimir effect in non-abelian gauge theories, we highlight its importance in quantum field theory. For example, in quantum chromodynamics, the Casimir effect has been pivotal in building our understanding of the phenomenologically relevant MIT bag model \cite{Chodos:1974pn,DeGrand:1975cf,Chodos:1974je} of hadrons which describes hadron spectroscopy, confinement and other hadron properties. 

In this work, we study the Casimir effect in non-abelian gauge theories in (2+1)D and (3+1)D for the gauge group SU(3) using lattice techniques. Inspired by the first-principle numerical simulations of Ref.\ \cite{Chernodub:2018pmt}, we apply chromoelectric boundary conditions on the lattice to formulate new as of yet unexplored types of geometries. We extend the results of the Casimir potential in (3+1)D gauge theories for parallel conducting plates recently computed in Ref.\ \cite{Chernodub:2023dok} to SU(2) and provide new results for the Casimir potential in (3+1)D for a symmetrical tube and box respectively. These geometries are shown in Fig.\ (\ref{fig:geometries}). The asymmetrical tube and box are studied in Ref.\ \cite{Ngwenya:2025mpo}.

Moreover, we explore the Casimir energy dependence on the temperature given that Yang-Mills theories in infinite volume exhibit a phase transition where the theory deconfines into a plasma of gluons. As such, we investigate the Polyakov loop (deconfinement order parameter) as we change temperatures. 
%In order to understand the gluodynamics of the restricted gluon modes inside the configuration volume, we compare the Polyakov loop of the fields on the interior and exterior (vacuum) of each configuration.

\section{Boundaries and Electromagnetic Fields}
\label{chapter:boundaries_em_fields}
%The local expectation of the energy density of the gluon fields,
%\begin{eqnarray}
%    T^{00} = \frac{1}{2}(\Vec{E^2} + \Vec{B^2}),
%\end{eqnarray}
%describes the energy of the vacuum fluctuations in the field
The energy-momentum tensor associated with the Yang-Mills Lagrangian in Minkowski space,
\begin{eqnarray}
    \mathcal{L}_{YM} = -\frac{1}{4} F^a_{\mu\nu} F^{\mu\nu,a} ,
\end{eqnarray}
is given by,
\begin{eqnarray}
    \label{eqn:field_strength}
    T^{\mu\nu} = -F^{\mu\alpha,a} F^{\mu,a}_{\alpha} + \frac{1}{4} \eta^{\mu\nu} F^{a}_{\alpha\beta} F^{\alpha\beta,a},
\end{eqnarray}
where $F_{0i} = E_i$, $F_{ij} = \epsilon_{ijk}B_k$ and $\frac{1}{4}F_{ij}F_{ij} = \frac{1}{2}B_iB_i$. We work on a Euclidean lattice using the coordinate system notation, $(x,y,z,t)$, thus need to transform the field-strength tensor in Eq.\ (\ref{eqn:field_strength}) into this new coordinate system. The resulting energy density in Euclidean space whose local expectation value describes the energy of vacuum fluctuations of the gluon field is,
\begin{eqnarray}
\label{eqn:energy_density_component}
    T^{00} = \frac{1}{2}(\Vec{E^2} - \Vec{B^2}).
\end{eqnarray}

%\subsection{Parallel Plates Fields}
In the parallel plate geometry where the two plates are placed at $x_0$ and $x_1$, separated by a distance $R$, one can explore the symmetries of the Euclidean field strength tensor by exploiting the rotational symmetries of the geometry in four-dimensional Euclidean space. The three possible rotations result in the following equivalence relations in the expectation values of the field components,
\begin{eqnarray}
    \label{eqn:F1}
    \text{F}1: \quad \langle B_z^2 \rangle = \langle B_y^2 \rangle \quad \text{and} \quad \langle E_z^2 \rangle = \langle E_y^2 \rangle\\
    \label{eqn:F2}
    \text{F}2: \quad \langle B_z^2 \rangle = \langle E_x^2 \rangle \quad \text{and} \quad \langle B_x^2 \rangle = \langle E_z^2 \rangle\\
    \text{F}3: \quad \langle B_y^2 \rangle = \langle E_x^2 \rangle \quad \text{and} \quad \langle B_x^2 \rangle = \langle E_y^2 \rangle
    \label{eqn:F3}
\end{eqnarray}
where $\text{F}1$ is obtained from the rotation about the $xt$ plane, $\text{F}2$ about the $xz$ plane and $\text{F}3$ about the $xy$ plane. 

The electromagnetic field strength tensor is related to the lattice \textit{plaquette} variable according to,
\begin{eqnarray}
    \label{eqn:plaquette_derivatives}
     U_{\mu\nu}(n)  &=& 1 + ia^2F_{\mu\nu}(n) - \frac{a^4}{2}\left(F_{\mu\nu}(n)\right)^2 + ...,
\end{eqnarray}
where $a$ is the lattice spacing, and the continuum definition of the field strength tensor generalised from the definition in electrodynamics has been used and the forward finite differences method is employed for the derivative terms. On the lattice, the plaquette is represented by the product of four link variables forming a square,
\begin{eqnarray}
     U_{\mu\nu}(n) &=&  U_{\mu}(n)U_{\nu}(n+\hat{\mu}) \nonumber\\
     & &\times U_{-\hat{\mu}}(n +\hat{\mu}+\hat{\nu})U_{-\hat{\nu}}(n+\hat{\nu}),\\
                    &=&  U_{\mu}(n)U_{\nu}(n+\hat{\mu})U_{\mu}(n+\hat{\nu})^{\dag}U_{\nu}(n)^{\dag}.
                    \label{eqn:plaquette}
\end{eqnarray}

The sum of all possible plaquettes on the lattice describes the Wilson gauge action, with each plaquette counted with a single orientation. The Wilson action is given by,
\begin{eqnarray}
    \label{eqn:wilsonaction}
    S \left[ U\right] \equiv \frac{\beta}{N_c} \sum\limits_{n\in\Lambda} \sum\limits_{\mu<\nu} \text{Re Tr} \left[ 1 -  U_{\mu\nu}(n) \right],
\end{eqnarray}
where the sum is performed over all lattice points containing a plaquette and the Lorentz indices. The pre-factor, $\beta$ is the inverse coupling,
\begin{eqnarray}
    \beta = \frac{2N_c}{g^2},
\end{eqnarray}
and provides the physical scale on the lattice. The electric-type boundary conditions are enforced by the following condition \cite{Chernodub:2018pmt};
\begin{equation}
%\[
    \beta_P = 
\begin{dcases}
    \beta, & P\notin \mathcal{S}\\
    \lambda_w \beta,              & P\in \mathcal{S},
\end{dcases}
%\]
\label{eqn:boundary_condition}
\end{equation}
requiring the coupling at the surface of the perfect conductor material (given by the limit, $\lambda_w \to \infty$) to be different from the vacuum inverse coupling, where the subscript $P$ refers to plaquette and $\mathcal{S}$ refers to the worldvolume of the conductor material. We define $S^{(k)}_{P_{ij}}$ as the action of the $k$-th configuration with plaquettes $P_{ij}$ lying on the worldvolume of the boundary.

\begin{figure}[!htbp]
    \centering
    \subfigure[Distance $R_{1a}=1$\label{subfig:fields_3DSU3Plates_R1}]{{\includegraphics[width=0.9\linewidth]{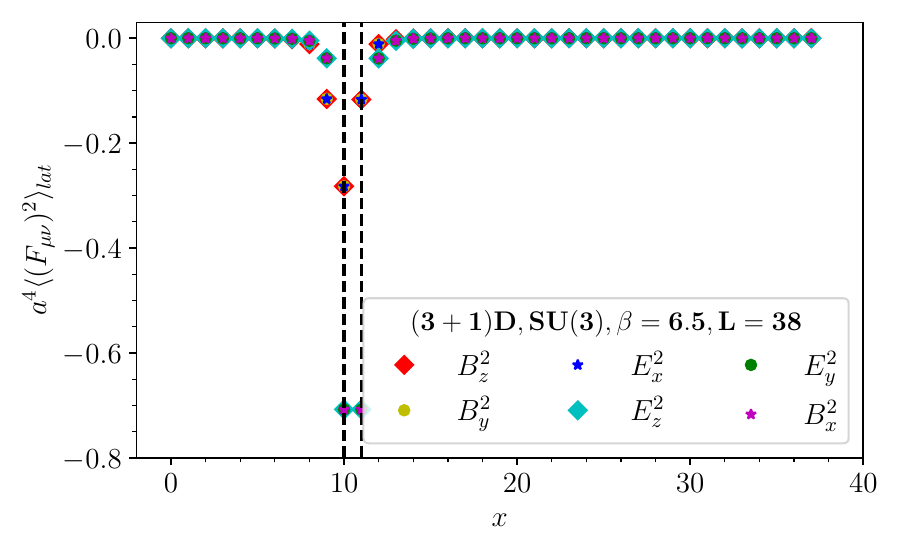} }}%
    \hspace{1em}
    \subfigure[Distance $R_{1a}=2$\label{subfig:fields_3DSU3Plates_R2}]{{\includegraphics[width=0.9\linewidth]{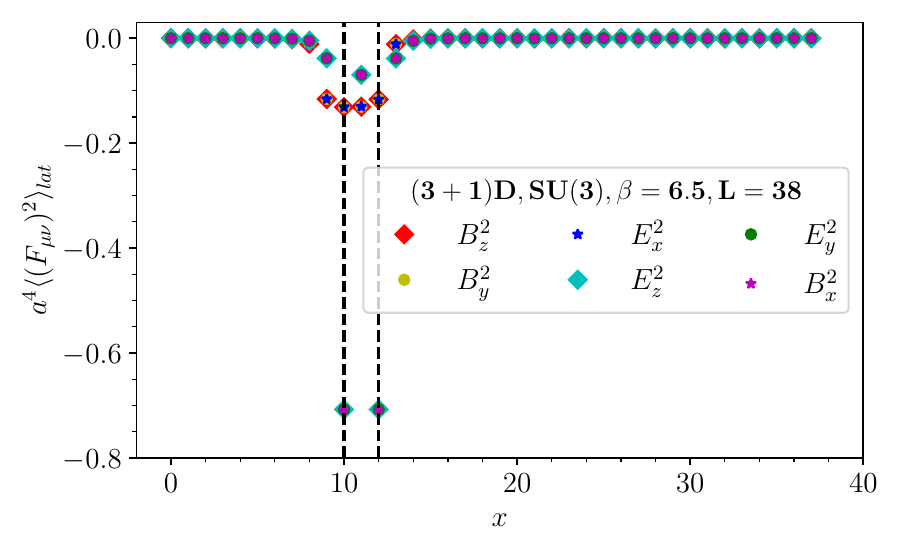} }}
    \hspace{1em}
    \subfigure[Distance $R_{1a}=10$\label{subfig:fields_3DSU3Plates_R10}]{{\includegraphics[width=0.9\linewidth]{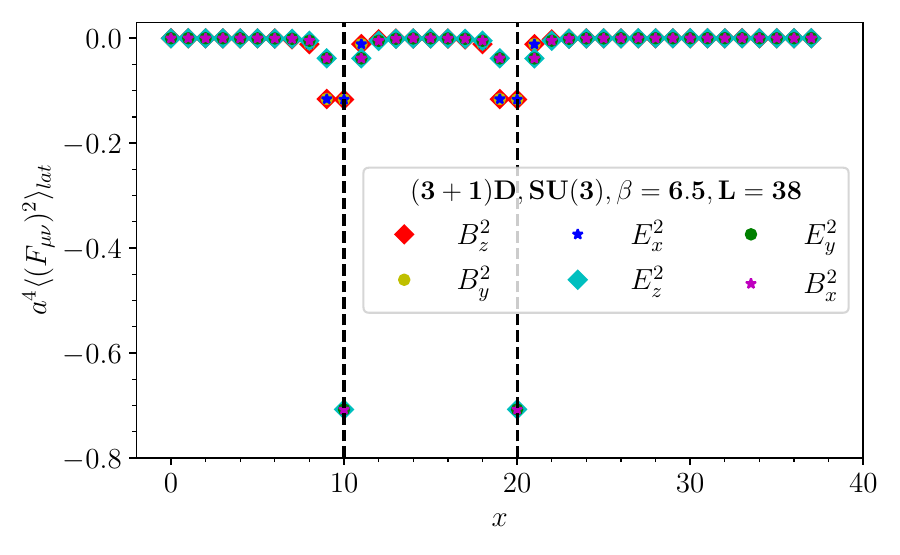} }}
    %\hspace{1em}
    \caption{Expectation values of the squared field strength tensor components in $(3+1)$D SU(3) along the $x$-axis orthogonal to the plates (represented by dashed lines) placed a distance $R$ apart.}%
    \label{fig:fields_3DSU3Plates}%
\end{figure}

In the absence of the plates, the vacuum fields components in Eq.\ (\ref{eqn:energy_density_component}) are indistinguishable and fluctuate around a common expectation value as expected for a zero temperature lattice. We show the numerical results of the field expectation values in the presence of plates at varying separation distances normalised by subtracting the vacuum field expectation values in Fig.\ (\ref{fig:fields_3DSU3Plates}). The field components equivalence relations provided in Eq.\ (\ref{eqn:F1} - \ref{eqn:F3}) can be seen directly from all the sub-figures, where the error-bars are omitted since they are small. We observe that field components on the worldvolume of the plates, $E_z$, $E_y$ and $B_x$ increase in magnitude as the distance to the plates is decreased. We also observe that the remaining field components also react to the presence of the plates and are slightly chipped. This behaviour is expected from the structure of the local action, where updating a single link variable affects the six plaquettes attached to it.

Note that there is a clear symmetry in the orientation of the equivalent field fluctuations at the position of the plates because these are artificially modified through the boundary condition. This is accompanied by an asymmetry in the fields around the plates as a consequence of the naive forward finite differences approach in taking the derivatives in the Wilson field-strength formulation given in Eq.\ (\ref{eqn:plaquette_derivatives}). This topic is subject to ongoing research within the lattice community, see for example Ref.\ \cite{Rothkopf:2021jye} for an expansive discussion.

\begin{figure}[!htb]
    \centering
    \subfigure[Top-view\label{subfig:fields_tube_all1}]{{\includegraphics[width=0.9\linewidth]{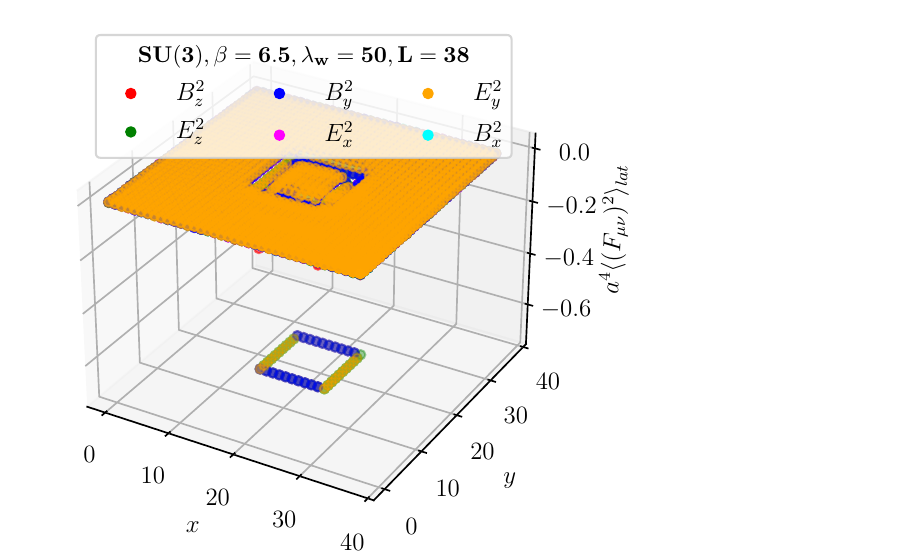} }}%
    \hspace{1em}
    \subfigure[Side-view\label{subfig:fields_tube_all2}]{{\includegraphics[width=0.9\linewidth]{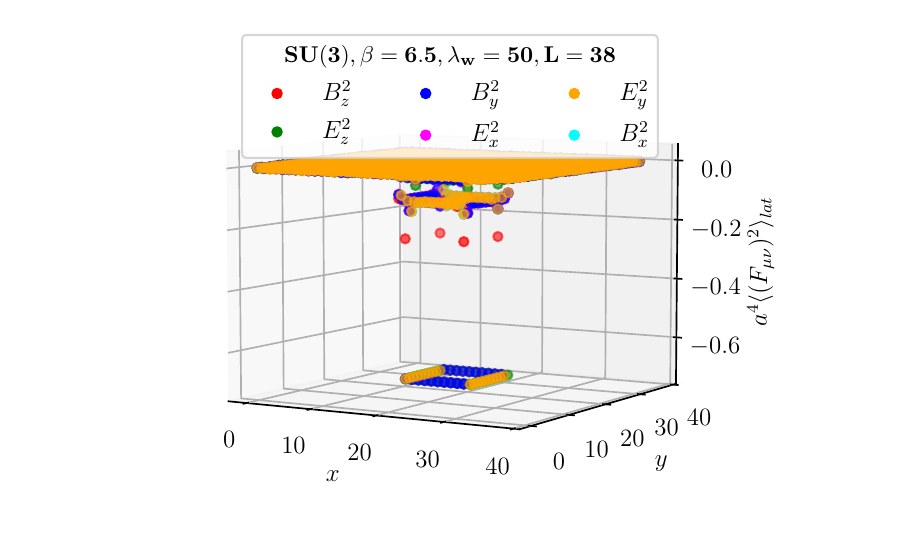} }}
    %\hspace{1em}
    \caption{Three dimensional expectation values of the squared field strength tensor components in $(3+1)$D SU(3) for a symmetrical tube with side lengths $R_{1a}=10$ and faces at $x=y=14a$ and $x=y=24a$.}%
    \label{fig:fields3D_tube_all}%
\end{figure}

%\subsection{Symmetric Tube Fields}
In the case of the symmetrical tube, exploiting rotational symmetries of the geometry results in the F3 field equivalence relation given in Eq.\ (\ref{eqn:F3}). We show a three-dimensional visualisation of the field configurations for a symmetrical tube with side lengths $R_{x_{1a}} = R_{y_{1a}} =10$ in Fig.\ (\ref{fig:fields3D_tube_all}). The elongated $\hat{z}$ direction has been integrated out since the fields should be the same along this direction. In Fig.\ (\ref{subfig:fields_tube_all1}) we observe that all field components fluctuate around zero in vacuum, far from where the tube is positioned. As we approach the tube (from both the outside or the inside), the surrounding plaquettes start experiencing the gauge field contributions from the boundaries of the tube and the fields change. This is seen in Fig.\ (\ref{subfig:fields_tube_all2}) where the magnitude of change increases the closer one is to the face of the tube. We also observe a relatively large change in the magnitude of the fields at the four corners of the tube, followed by maximal change on the faces of the tube. Numerical evidence of the F3 field equivalence relations is provided in Ref.\ \cite{Ngwenya:2025mpo}.

\begin{figure}[!htb]
    \centering
    \subfigure[$z=10a$\label{subfig:fields_box_z10}]{{\includegraphics[width=0.9\linewidth]{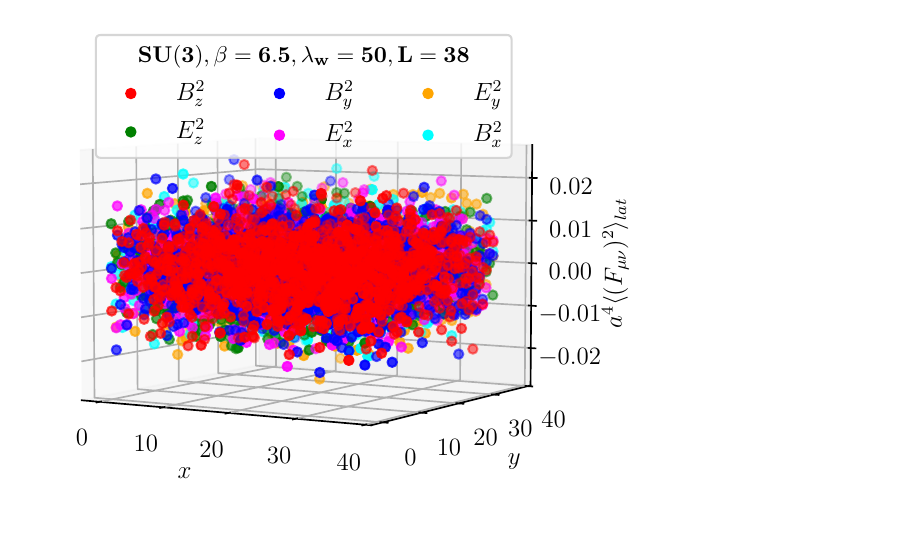} }}%
    \hspace{1em}
    \subfigure[$z=13a$\label{subfig:fields_box_z13}]{{\includegraphics[width=0.9\linewidth]{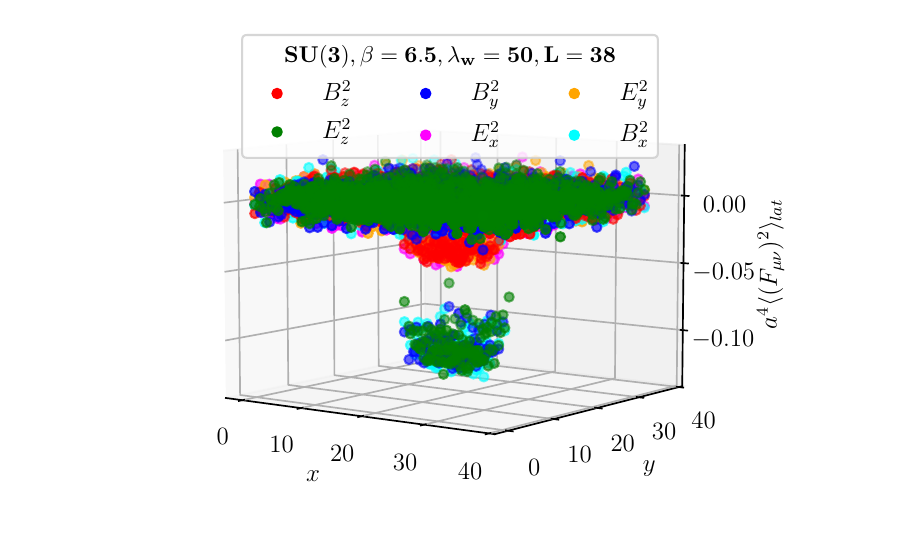} }}
    %\hspace{1em}
    \caption{Three dimensional expectation values of the squared field strength tensor components in $(3+1)$D SU(3) outside a symmetrical box with side lengths $R_{1a}=10$ and faces at $x=y=z=14a$ and $x=y=z=24a$.}%
    \label{fig:fields3D_box_out}%
\end{figure}

Lastly, we consider the geometry of a symmetrical box where all rotational symmetries are broken because we have fixed the spatial 3D axis, leaving only the temporal direction free. In Fig.\ (\ref{subfig:fields_box_z10}), we show the vacuum field expectation values at a distance $R_{1a}=4$ away from the box. The surrounding gauge fields retain their vacuum expectation values and fluctuate around zero because they are vacuum normalised. Far enough outside the box, the surrounding plaquettes do not experience any plaquette contributions from the boundaries of the box. The various electromagnetic field components are indistinguishable.

\begin{figure}[!htb]
    \centering
    \subfigure[$z=14a$ Top-view\label{subfig:fields_box_all1}]{{\includegraphics[width=0.9\linewidth]{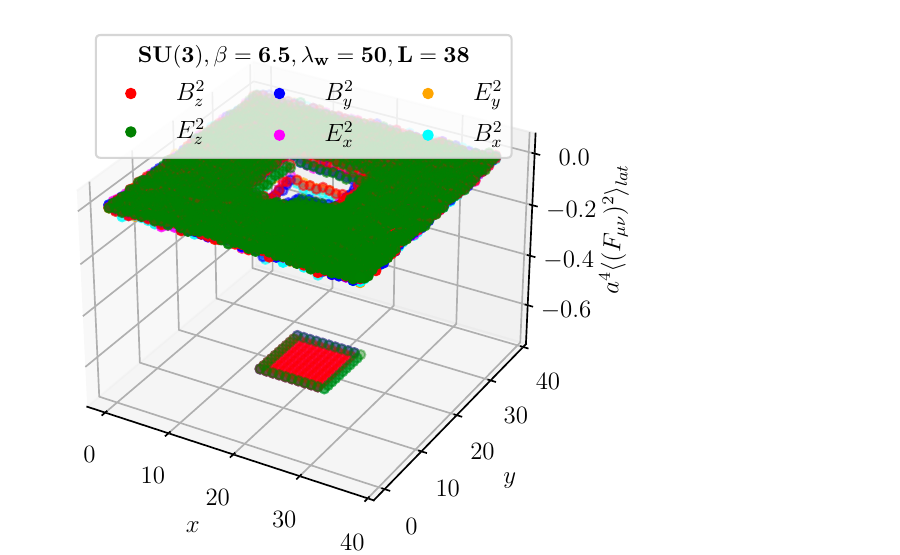} }}%
    \hspace{1em}
    \subfigure[$z=14a$ Side-view\label{subfig:fields_box_all2}]{{\includegraphics[width=0.9\linewidth]{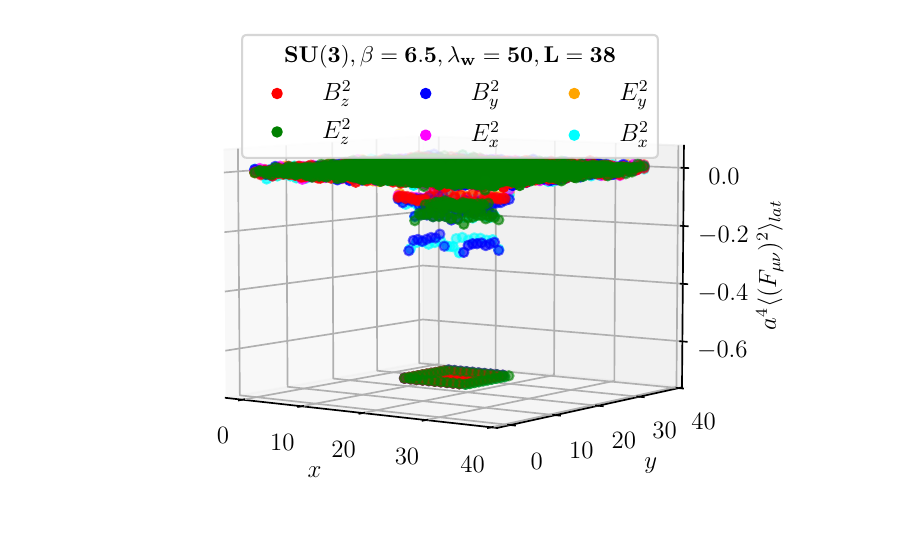} }}
    %\hspace{1em}
    \caption{Three dimensional expectation values of the squared field strength tensor components in $(3+1)$D SU(3) on the face (at $z=14a$) of a symmetrical box with side lengths $R_{1a}=10$ and faces at $x=y=z=14a$ and $x=y=z=24a$.}%
    \label{fig:fields3D_box_face}%
\end{figure}

As one approaches one of the faces of the box from the outside, we start seeing a change in some of the field components at varying degrees as shown in Fig.\ (\ref{subfig:fields_box_z13}) a distance $R_{1a}=1$ away from the lower face of the box. The intensity of change in magnitude of the individual electromagnetic field components is dependent on the position around the box (i.e., different components change at varying intensities around the different faces depending on the contributing plaquettes). In this region, the vacuum plaquettes are now attached to link variables that are in contact with the boundaries of the box.

In Fig.\ (\ref{fig:fields3D_box_face}), we show a cross section of the electromagnetic field components on the bottom face of the box at $z=14a$. We observe a hollow region in the field strength indicating that all field components have the largest magnitude on the face. Similarly to the tube geometry, the field components with largest magnitude depend on the face under consideration. In this case, we are looking at a face in the $xy$-plane, therefore the $E_x$, $E_y$ and $B_z$ experience the largest magnitude on this face. The field components on cross sections inside the box are similar to those of the tube, including the equivalence of the fields described by the F3 symmetry. We refer the reader to Ref.\ \cite{Ngwenya:2025mpo} for a comprehensive discussion on these field contributions.

\section{\label{results}Results}

In this section, we present results of the non-abelian Casimir effect in SU(3) for the geometry of parallel plates as well as a symmetrical tube and box respectively. The geometries of an asymmetrical tube and asymmetrical box are presented in Ref.\ \cite{Ngwenya:2025mpo}. Our calculations are performed on a periodic lattice volume with chromoelectric boundary conditions on the walls of our geometries. In the geometries of the tube and box where the plate-size expands with increasing separation distance, we discuss techniques that can be used to correctly account for the energy from creating the boundaries. We show that the resulting Casimir potential is attractive for all the geometries considered.

We use the naive Wilson action with the Metropolis algorithm to generate the field configurations. We take a minimum of $5\times 10^3$ thermalisation steps and compute the integrated autocorrelation time, which is then used to discard correlated measurements in the Markov chain and ensures that we only use statistically independent measurements. Our errors are computed using the Jackknife method. A continuum extrapolation is postponed to future study.

%%%%%%%%%%%%%%%%%%%%%%%%%%%%%%%%%%%%%%%%%%%%%%%%%%%%%%%%%%%%%%%%%%%%%%%%%%%%%%%%%%%%%%%%%%%%%%%%%%%%%%%%%%%%%%%%%%%%%%
\subsection{The Casimir Potential}

\subsubsection{Parallel Plates}
The energy density for the plates can assume three equivalent forms due to the field equivalence relations obtained from rotational symmetry discussed in the preceding chapter,
\begin{eqnarray}
     \varepsilon(x)_{\text{plates}} &=& \frac{1}{2} \sum\limits_i \left[ \langle B_i^2 \rangle - \langle E_i^2 \rangle \right], \quad i=x,y,z\label{eqn:plates_Edensity_F0}\\
     &=& \frac{1}{2}\left[ \langle B_y^2 \rangle - \langle E_y^2 \rangle \right] \label{eqn:plates_Edensity_F2} \\
     &=& \frac{1}{2}\left[ \langle B_z^2 \rangle - \langle E_z^2 \rangle \right] \label{eqn:plates_Edensity_F3},
\end{eqnarray}
where we have used the F$2$ symmetry in Eq.\ (\ref{eqn:F2}) to obtain Eq.\ (\ref{eqn:plates_Edensity_F2}) and F$3$ symmetry in Eq.\ (\ref{eqn:F3}) to obtain Eq.\ (\ref{eqn:plates_Edensity_F3}).

The lattice expression for the energy density reads,
\begin{eqnarray}
     \varepsilon(x)_{\text{plates}}^{\text{lat}} &=& \langle S_{P_{ij}} \rangle = \frac{1}{N_{\tau}} \sum\limits_{N_{y}} \sum\limits_{N_{z}} \sum\limits_{N_{\tau}} S_{P_{ij}},
     \label{eqn:Edensity_plates_lat}
\end{eqnarray}
where $P_{ij}$ represents the plaquettes which lie on the worldvolume of the plates. We have checked that the definitions based on Eq.\ (\ref{eqn:plates_Edensity_F0} - \ref{eqn:plates_Edensity_F3}) are equivalent in our simulations. See Ref.\ \cite{Ngwenya:2025mpo} for the potential computed using the different symmetry relations. We use $S_{P_{ij}}$ for shorthand, representing the average of $S^{(k)}_{P_{ij}}$ over all $k$-configurations.

Note that the energy density has no direct dependence on the individual plaquette components, $P_{ij}$ contributing to the total energy density due to the summation. We represent the energy density, $\langle S_{P_{ij}} \rangle$ with the indices to signal that only contributions from selected plaquettes are considered and for consistency with the notation in the available literature. We obtain the three-dimensional normalised Casimir energy by integrating over the normal direction and applying the normalisation condition to remove the energy contribution from the plates,
\begin{eqnarray}
    V^{\text{lat}}_{\text{Cas}}(R) &=& \left[ \int_{dx}\, \varepsilon(x)_{\text{plates}} \right]_{R-R_0} = \left[ \sum\limits_{N_{x}} \varepsilon(x)_{\text{plates}}^{\text{lat}} \right]_{R-R_0}\\
    &=&  \sum\limits_{N_{x}} \left[ \langle S_{P_{ij}} \rangle_R - \langle S_{P_{ij}} \rangle_{R_0} \right] = \langle \langle S_{P_{ij}} \rangle \rangle,
    \label{eqn:3d_casimir_lattice}
\end{eqnarray}
given in lattice units.  On the lattice, the Casimir potential is equivalent to the normalised expectation value of the total energy of the system,
\begin{eqnarray}
    \langle \langle S_{P_{ij}} \rangle \rangle &=& \left[ \frac{1}{N_{\tau}} \sum\limits_{N_{x}} \sum\limits_{N_{y}} \sum\limits_{N_{z}} \sum\limits_{N_{\tau}}  S_{P_{ij}} \right]_{R-R_0},
    \label{eqn:3d_action_lattice}
\end{eqnarray}
and we highlight this definition for reasons that will become clear in the subsequent subsections.

Due to periodic boundary conditions, the plates extend infinitely in the $\hat{y}$ and $\hat{z}$ directions. Since we consider the energy associated with a single lattice volume, we take as plate area, $A=N_z\times N_y$ which can be combined with Eq.\ (\ref{eqn:Edensity_plates_lat}) to find an expression for the energy per unit area of the plates. The physical potential scales as $V^{\text{phys}}_{\text{Cas}} =  V^{\text{lat}}_{\text{Cas}}/a$, giving a well-defined expression for the total energy through an extension of the two-dimensional case proposed in Ref.\ \cite{Chernodub:2018pmt} with the tadpole improved coupling $\beta_I$,
\begin{eqnarray}
    V_{\text{Cas}}(R_{\text{phys}})/\sigma &=& \frac{1}{a\sqrt{\sigma}} \left( \frac{\beta_I}{\beta} \right)^4 \langle \langle S_{P_{ij}} \rangle \rangle,
    \label{eqn:Vcas3D_total_scaled}
\end{eqnarray}
where $\sigma$ is the string tension and the Casimir energy per unit area of the plates scales as $V^{\text{phys}}_{\text{Cas}} =  V^{\text{lat}}_{\text{Cas}}/a^3$, and the resulting potential per unit area is,
\begin{eqnarray}
    V_{\text{Cas}}(R_{\text{phys}})/\sqrt{\sigma^3} &=& \frac{1}{a^3 \sqrt{\sigma^3}} \left( \frac{\beta_I}{\beta} \right)^4 \langle \langle S_{P_{ij}} \rangle \rangle,
    \label{eqn:Vcas3D_formula_scaled}
\end{eqnarray}
and the string tension is used to express the potential in dimensionless units.

The normalisation is enforced by subtracting the lattice action expectation value with the parallel plates placed at separation distance, $R_0\to \infty$ apart, which on our finite periodic lattice we choose to be at half the lattice spatial extent, $R_0 = L/2$. Essentially, we compute the total energy of the system with plates placed a finite distance, $R$ apart, and from this, we subtract the total energy of the system with the plates placed as far apart as possible. At large enough distances, the Casimir interaction will diminish leaving only the vacuum energy contributions and the energy of creating the boundaries. We express this as,
\begin{eqnarray}
    V^{\text{lat}}_{\text{Cas}}(R) &=& E^{\text{Tot}}_R - E^{\text{Tot}}_{R_0}\\
    &=& [E_{\text{Cas}}(R) + E_{\text{Plates}} + E_{\text{Vac}}]_R \nonumber \\ && - [E_{\text{Plates}} + E_{\text{Vac}}]_{R_0},
    \label{eqn:plates_energy_normalisation}
\end{eqnarray}
where the zero-point energy is independent of $R$, but so is the energy contribution from the plates because the size of the plates is fixed. We will return to the discussion of the energy contributions from imposing boundary conditions in the subsequent sections where it contributes a non-constant value. The normalisation condition takes care of two energy contributions to the system, the first is the cancellation of the vacuum ultraviolet divergences such that the energy density describes a local finite quantity. The second is the energy contribution from imposing the chromoelectric boundary condition (geometry of two parallel plates) on the QCD vacuum, i.e., the act of putting the plates modifies the zero-point energy.

\begin{figure}[!htbp]
    \centering
    \subfigure[SU(2)\label{subfig:casimir_plates_3dsu2}]{{\includegraphics[width=0.9\linewidth]{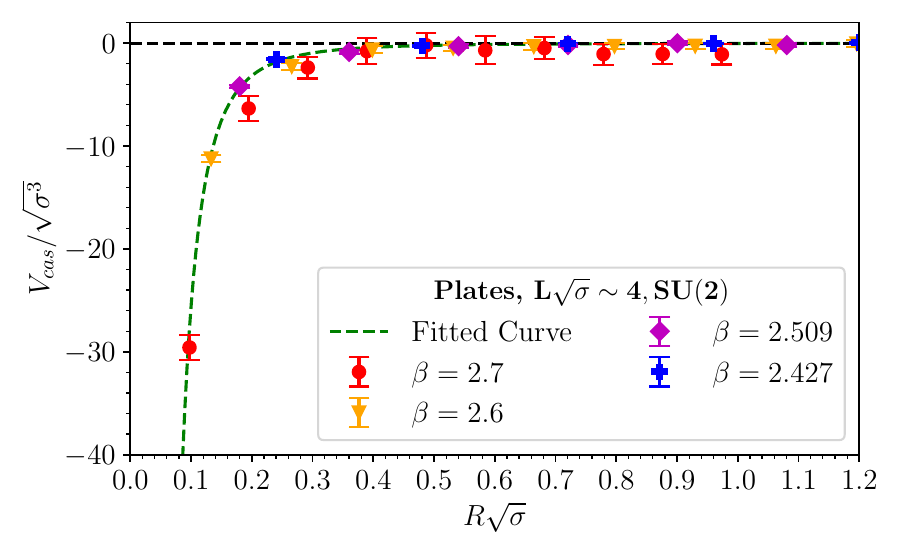} }}%
    \hspace{1em}
    \subfigure[SU(3)\label{subfig:casimir_plates_3dsu3}]{{\includegraphics[width=0.9\linewidth]{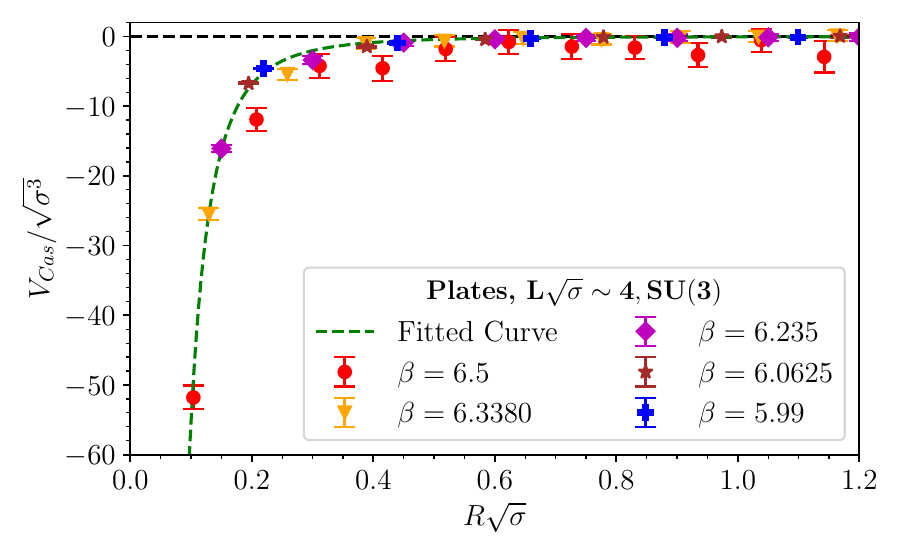} }}
    \hspace{1em}
    \subfigure[SU(3)/SU(2) Ratio\label{subfig:casimir_plates_ratio}]{{\includegraphics[width=0.9\linewidth]{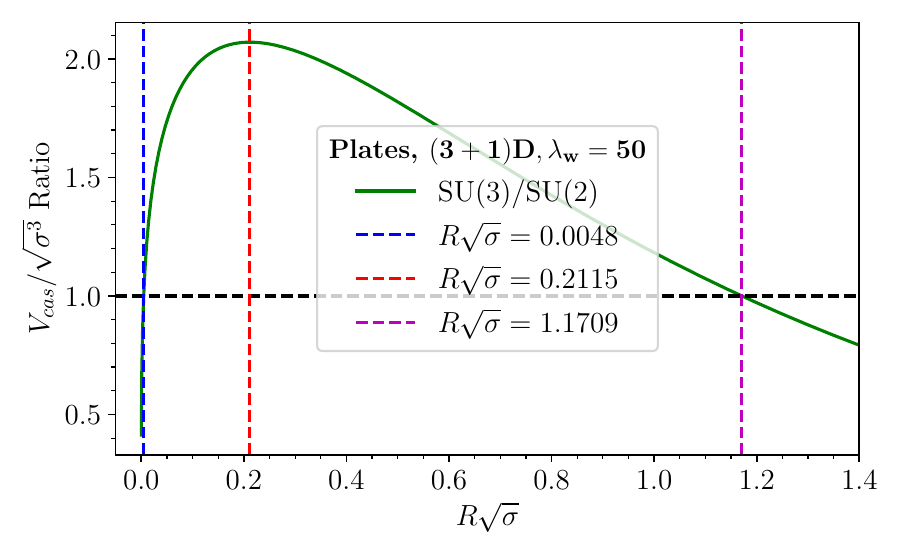} }}
    %\hspace{1em}
    %\hspace{1em}
    \caption{The Casimir potential per unit area between two parallel plates separated by a distance $R\sqrt{\sigma}$.}
    \label{fig:casimir_plates}
\end{figure}

The dimensionless physical Casimir potential per unit area of the plates is shown in Fig.\ (\ref{subfig:casimir_plates_3dsu2}) in SU(2) and Fig.\ (\ref{subfig:casimir_plates_3dsu3}) in SU(3) in units of the string tension. The potential is negative, its functional form indicates that the Casimir force, described by the slope of the potential, experienced by the plates in the non-abelian gauge theory is attractive. The fitting function that we have employed for the potential of the plates has a similar functional form and same physical interpretation as the one used for the wires in (2+1)D \cite{Chernodub:2018pmt}. It is given by,
\begin{equation}
    V_{\text{Cas}}^{\text{fit}} (R) = -(N_c^2-1)\frac{\pi^2}{1440} \frac{1}{R^{(\nu+3)}\sigma^{(\nu+3)/2}} e^{-M_{\text{Cas}}R},
    \label{eqn:vcas_fit_plates}
\end{equation}
where the power of $\sigma$ is chosen to give a dimensionless potential, i.e., $V_{\text{Cas}}/\sqrt{\sigma^3}$. The coefficients are obtained from the expected tree level behaviour of the potential for a non-interacting theory, $M_{\text{Cas}} = \nu =0$ \cite{Ambjorn:1981xw}, 
\begin{equation}
    V_{\text{Cas}}^{\text{tree}} (R) = -(N_c^2-1)\frac{\pi^2}{1440} \frac{1}{R^{3}},
    \label{eqn:vcas_tree_plates}
\end{equation}
describing $(N_c^2-1)$ additive contributions from the non-interacting gluon fields. 

Using the fitting functions, we compute the ratio of the potential in Fig.\ (\ref{subfig:casimir_plates_ratio}) to determine the dependence on the number of degrees of freedom. As we move from $N_c:2\to3$, at intermediate separation distances, the Casimir pressure experienced by the plates is stronger in SU(3), peaking at approximately twice as strong at $R\sqrt{\sigma} \sim 0.2$. At very small and at large separation distances where we expect a negligible Casimir force, the pressure is stronger in SU(2). While we do not provide a physical interpretation at this point, we state these results as they point out a need for further investigation of the Casimir effect and relevant fitting functional form between the two theories.

The Casimir potential for the plates has also been recently computed in Ref.\ \cite{Chernodub:2023dok} where a comprehensive discussion is given for the presence of a boundary induced deconfined phase transition in the region between the plates. We will expand on this discussion when we look at the temperature dependence of the potential in the next section. The SU(3) Casimir potential that we obtain has similar qualitative features to the potential obtained in Ref.\ \cite{Chernodub:2023dok}, however the rate of decay of the potential varies particularly at short separation distances where our measured potential decays more rapidly. In addition, the Casimir mass obtained by Chernodub et al.\, $M_{\text{Cas}} = 1.0(1)\sqrt{\sigma}$ differs substantially from the mass obtained in our analysis \cite{Ngwenya:2025mpo}, likely due to the fitting functional form. We highlight this difference in the measured potential and Casimir mass as an aspect that needs to be investigated further in future studies of the parallel plate geometry.

\subsubsection{Symmetrical Tube}
The next geometry that we consider in the non-abelian theory is a hollow symmetrical tube. In the preceding section, we showed that there exists an equivalence of the $\hat{x}$ and $\hat{y}$ field components according to the F3-symmetry in Eq.\ (\ref{eqn:F3}) such that the expression for the energy density of the system in Euclidean space reduces to only the $\hat{z}$ components,
\begin{eqnarray}
    \varepsilon_{\text{tube}}(x,y) &=& \frac{1}{2}\left[ \langle B_z^2 \rangle - \langle E_z^2 \rangle \right]  \label{eqn:tube_Edensity},
\end{eqnarray}
which depends on both $x$ and $y$ because the gluon fields are now enclosed in a hollow cavity and the Casimir force acts in two directions.

On the lattice, the energy density is given by a sum over the plaquettes on the worldvolume of the tube,
\begin{eqnarray}
     \varepsilon(x,y)_{\text{tube}}^{\text{lat}} &=& \langle S_{P_{ij}} \rangle = \frac{1}{N_{\tau}} \sum\limits_{N_{z}} \sum\limits_{N_{\tau}} S_{P_{ij}},
\end{eqnarray}
and these plaquettes contributing to the action are reduced to $P_{xy}$ and $P_{zt}$ according to Eq.\ (\ref{eqn:tube_Edensity}). The expression for the Casimir potential of the tube reads,
\begin{eqnarray}
    V^{\text{lat}}_{\text{Cas}}(R) &=& \left[ \int_{d\mathcal{A}}\, \varepsilon(x,y)_{\text{tube}} \right]_{R-R_0} \\&=& \left[ \sum\limits_{N_{x}, N_{y}} \varepsilon(x,y)_{\text{tube}}^{\text{lat}} \right]_{R-R_0} = \langle \langle S_{P_{ij}} \rangle \rangle\\
    &=&  \sum\limits_{N_{x}, N_{y}} \left[ \langle S_{P_{ij}} \rangle_R - \langle S_{P_{ij}} \rangle_{R_0} \right],
    \label{eqn:tube_casimir_lattice}
\end{eqnarray}
where $\langle \langle S_{P_{ij}} \rangle \rangle$ is given by Eq.\ (\ref{eqn:3d_action_lattice}).

In the previous section, i.e.\ the geometry of parallel plates, the normalisation condition involved a simple constant since the area of the plates did not change with $R$. It is shown in Ref.\ \cite{Ngwenya:2025mpo} that the energy contribution from the plates was independent of the separation distance. However, the energy contribution from the tube depends on the separation distance $R$, because as $R$ increases, the size of the four rectangular faces forming the tube also increases and their energy contribution changes. We devise two approaches to account for this contribution.

\begin{figure}[!htbp]
    \centering
    \subfigure[Vacuum Subtracted Energy\label{subfig:tube_Svac_Slope}]{{\includegraphics[width=0.9\linewidth]{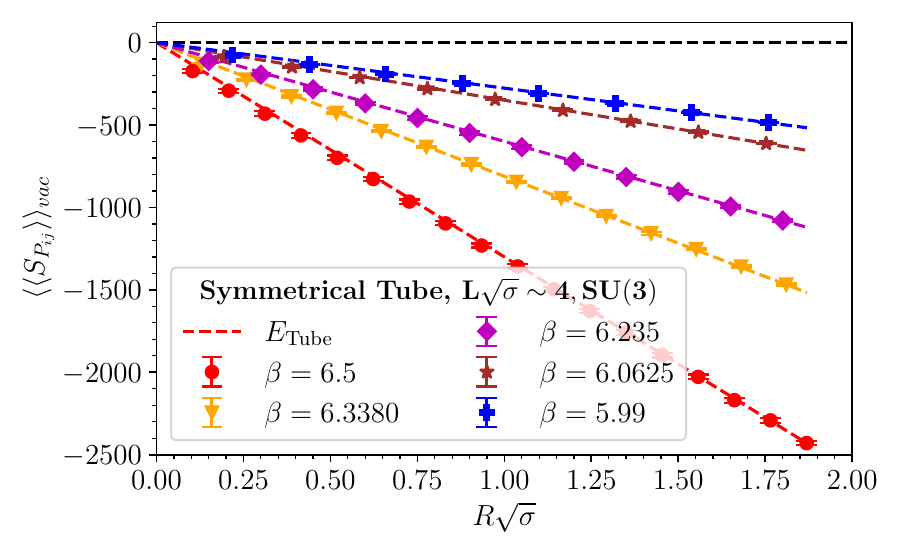} }}%
    \hspace{1em}
    \subfigure[$R_{\infty}$ Subtracted Energy\label{subfig:tube_Sinf_Slope}]{{\includegraphics[width=0.9\linewidth]{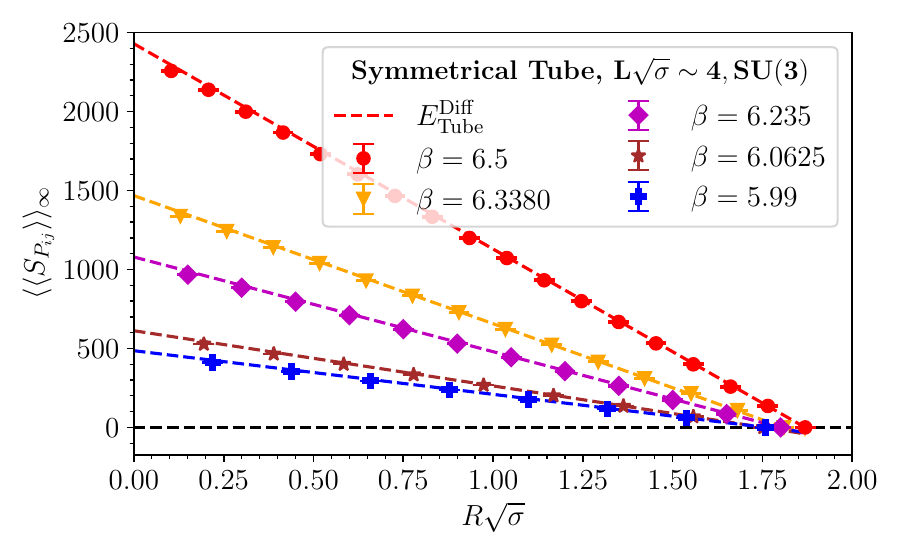} }}
    %\hspace{1em}
    \caption{Total energy of the system in a symmetrical tube in $(3+1)$D SU(3) for different couplings and different normalisation schemes in lattice units.}%
    \label{fig:tube_total_S_Slope}
\end{figure}

\begin{table}[!htbp]
\begin{ruledtabular}
    \centering
        %{\rowcolors{2}{green!80!yellow!50}{green!70!yellow!40}
        \begin{tabular}{cccc}
        %\hline
        %\multicolumn{3}{|c|}{Country List} \\
        %\hline
        Energy & $\mathbf{\beta}$ & $m$ & $c$\\
        \hline
         & 6.5 & -1299.96 & 2429.55  \\
         & 6.3380 & -811.02 & 1467.88\\
        $\langle \langle S_{P_{ij}} \rangle \rangle_{\infty}$ & 6.235 & -599.38 & 1079.10 \\
         & 6.0625 & -349.36 & 612.25 \\
         & 5.99 & -276.24 & 485.79\\
        %\hline
        \end{tabular}%}
    \caption{Linear fit parameters for the energy contributions from the boundaries of a symmetrical tube in SU(3) for lattice size, $L\sqrt{\sigma}\sim 4$.}
    \label{tab:tube_fit_params}
\end{ruledtabular}
\end{table}

In order to capture the tube's energy dependence on the separation distance, $R$ and consequently the size of the plates forming the tube, we plot the total energy of a symmetrical tube in Fig.\ (\ref{fig:tube_total_S_Slope}) using different normalisation schemes. In Fig.\ (\ref{subfig:tube_Svac_Slope}) we plot the vacuum subtracted total energy of the tube,
\begin{equation}
  E_{\text{Tot}}^{\text{Vac}} = E_{\text{Cas}} + E_{\text{Tube}}.
  \label{eqn:tube_vac_tot}
\end{equation} 
It is clear from observation that the total energy, varies linearly with separation distance. However, we also know from our findings for the plates, that at large separation distances, the Casimir energy contribution should be negligible.

\begin{figure}[!htbp]
\centering
\includegraphics[width=0.9\linewidth]{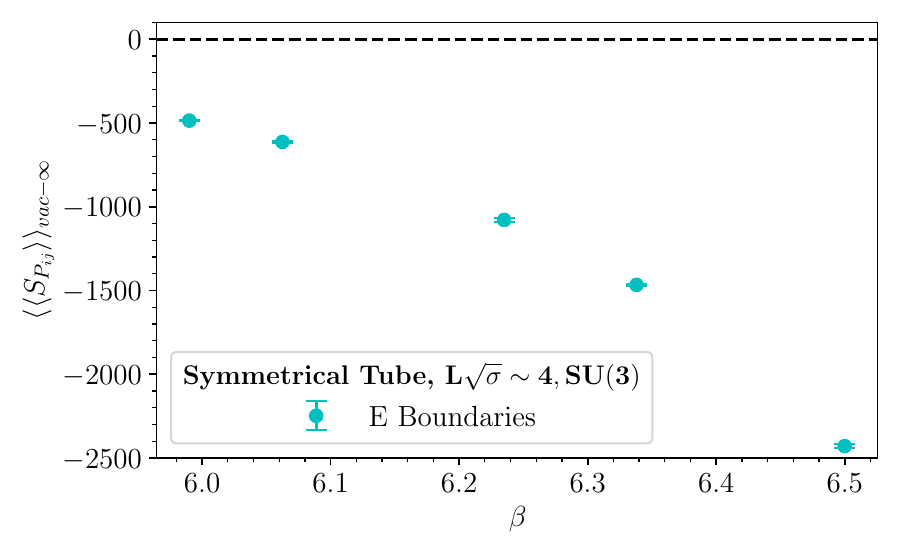}
\caption{The total energy required to create a symmetrical tube with side lengths $L\sqrt{\sigma}/2$ in $(3+1)$D SU(3) at varying lattice couplings.}
\label{fig:tube_Svac_Sinf}
\end{figure}

We therefore conclude that the energy contribution from creating the tube, $E_{\text{Tube}}$ increases linearly with increasing plate size, thus has the functional form,
%Consequently, we apply a linear fit to the total energy for $R\sqrt{\sigma} \gtrsim 1$ at different couplings,
\begin{equation}
    E_{\text{Tube}} = mR\sqrt{\sigma} + c,
    \label{eqn:linear_fit}
\end{equation}
with fit parameters, $m$ capturing the plate size dependence and $c$ providing the overall scaling. In the absence of the tube, i.e., at $R\sqrt{\sigma} =0$, the energy contribution from the tube vanishes. This allows us to set the constant term, $c=0$. At $R=L\sqrt{\sigma}/2$, which is the largest possible symmetric tube that we can create in the periodic lattice, the Casimir energy vanishes because there is an equal number of modes inside and outside the tube. We show the vacuum normalised total energy of the system at $R=L\sqrt{\sigma}/2$ in Fig.\ (\ref{fig:tube_Svac_Sinf}), which is equal to the energy of creating the large tube as per Eq.\ (\ref{eqn:tube_vac_tot}).

We use this result of the energy for the largest possible tube to fix the energy at the second endpoint of the linear function, and consequently calculate the slope of the resulting straight line which describes the energy from creating the tube. Subtracting this energy from the system's total energy in Eq.\ (\ref{eqn:tube_vac_tot}) isolates the Casimir potential. The resulting slope of the energy of the walls of a symmetrical tube shown in Fig.\ (\ref{fig:tube_total_S_Slope}) are provided in Table (\ref{tab:tube_fit_params}) at various gauge coupling constants, where $\langle \langle S_{P_{ij}} \rangle \rangle_{vac}$ has the same gradient, $m$ with $c=0$ for corresponding $\beta$.

\begin{figure}[!htbp]
    \centering
    \subfigure[Absolute Casimir Potential\label{subfig:tube_ETotVac_Slope}]{{\includegraphics[width=0.9\linewidth]{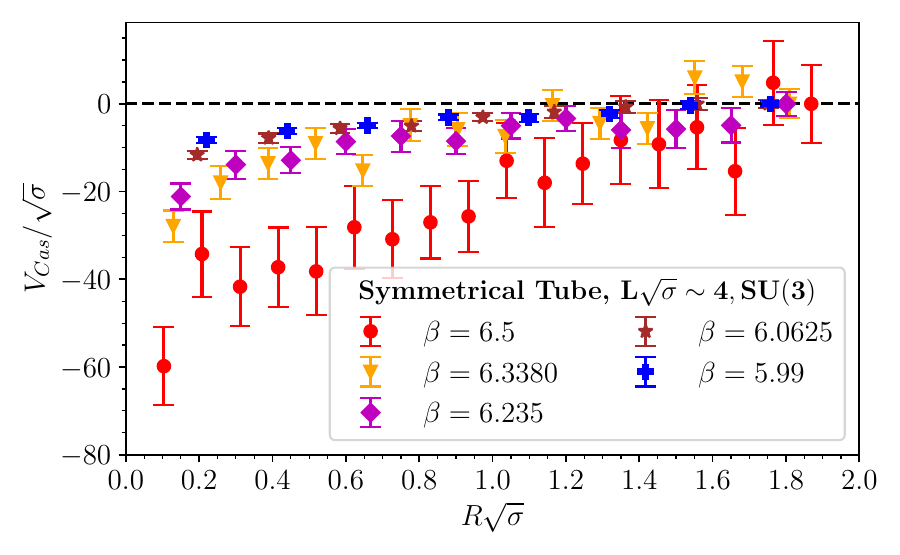} }}%
    \hspace{1em}
    \subfigure[Potential/Surface Area\label{subfig:tube_EVac_Slope}]{{\includegraphics[width=0.9\linewidth]{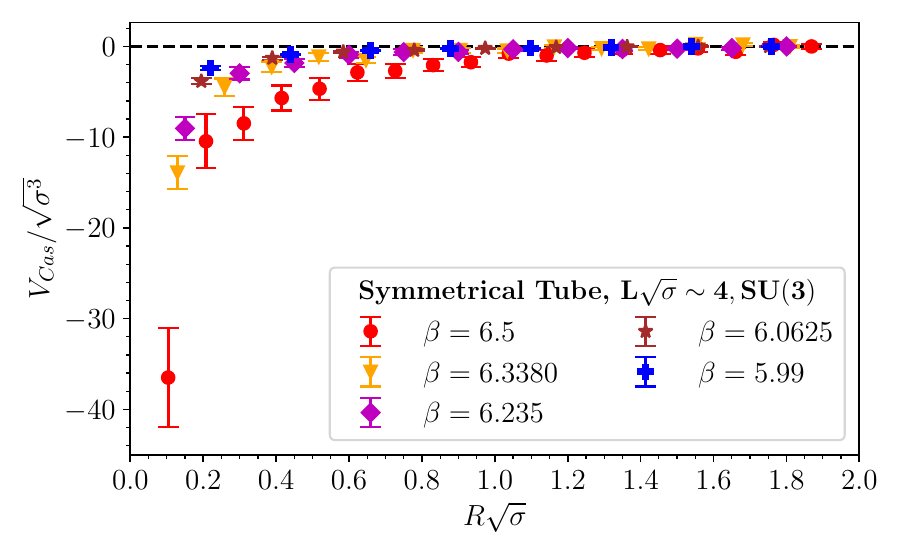} }}
    %\hspace{1em}
    \caption{The total Casimir potential for a symmetrical tube with side lengths $R\sqrt{\sigma}$ in SU(3).}
    \label{fig:Vcas_tube}
\end{figure}

In summary, the Casimir energy is isolated as follows;
\begin{enumerate}
    \item Compute the total energy of the system and apply the vacuum subtraction to remove UV divergences, remaining with the total energy in Eq.\ (\ref{eqn:tube_vac_tot}).
    \item Approximate the energy of the walls of the tube using a linear function. Take the constant term, $c=0$ because the energy from creating the boundaries vanishes in the absence of boundaries. Take the energy at $R=L\sqrt{\sigma}/2$ to be equal to the energy of creating the largest possible tube with side-length $R$ (where the Casimir energy vanishes). Draw a straight line between these two endpoints and calculate the slope. 
    \item Use the slope to perform a second subtraction on the total energy in Eq.\ (\ref{eqn:tube_vac_tot}) in order to remove the energy associated with creating the tube. This isolates the Casimir energy.
\end{enumerate}

The resulting physical absolute Casimir potential for the symmetrical tube as defined in Eq.\ (\ref{eqn:Vcas3D_total_scaled}) is shown in Fig.\ (\ref{subfig:tube_ETotVac_Slope}) and is attractive. While the Casimir energy per unit surface area of the symmetrical tube defined in Eq.\ (\ref{eqn:Vcas3D_formula_scaled}) is shown in Fig.\ (\ref{subfig:tube_EVac_Slope}). The corresponding total surface area of the symmetrical tube is, $A_{\text{lat}}=4\sigma RL$, where $R$ is the tube side-length and separation distance. Computing the potential per unit surface area introduces an additional dependence on the side-length of the tube, $R\sqrt{\sigma}$ which scales the diminishing potential with separation distance non-linearly.

In Fig.\ (\ref{subfig:tube_Sinf_Slope}), we show an alternative way of accounting for the energy contributions of the walls of the tube. Instead of performing a vacuum normalisation, we use the $R_{\infty}$ normalisation where we subtract from the total energy, the energy contribution with a large tube of side-length, $R=R_{\infty}=L\sqrt{\sigma}/2$. This is the same normalisation employed in the parallel plates geometry, except, the energy contribution from creating a larger tube of side-length, $R=R_{\infty}=L\sqrt{\sigma}/2$ is greater than the energy required to create a smaller tube with $R<L\sqrt{\sigma}/2$. Hence by subtracting the energy contribution at $R_{\infty}$, we eliminate the vacuum energy but remain with an energy difference between creating a small tube and a large tube according to,
\begin{eqnarray}
    E_{\text{Tot}}^{\infty} &=& E_{\text{Cas}} + E_{\text{Tube}}^{\text{Diff}} , \quad E_{\text{Tube}}^{\text{Diff}} = E_{\text{Tube}}^{\text{R}} -E_{\text{Tube}}^{\text{R}_{\infty}}
    \label{eqn:tube_Etot_infty}
\end{eqnarray}

Again, the resulting energy difference is linear and vanishes at $R=L\sqrt{\sigma}/2$ because the Casimir energy is negligible and the `inner tube' of side-length, $R$ has equal dimensions as the $R_{\infty}$ tube. Similarly to the vacuum normalisation, we employ a linear function to describe this energy difference and use it to renormalise the total energy in Eq.\ (\ref{eqn:tube_Etot_infty}), thus isolating the Casimir energy. This approach is discussed comprehensively in Ref.\ \cite{Ngwenya:2025mpo}, including a method of employing a linear fitting function for $R\gtrsim 1$ to obtain the energy from the walls of the tube.

We highlight the following important features of the symmetrical tube geometry based on our numerical results:
\begin{itemize}
    \item The Casimir potential is negative and has a positive slope, which implies that the resulting Casimir force is attractive. This result contrasts with the Casimir effect of a massless non-interacting scalar field computed in Ref.\ \cite{Mogliacci:2018oea} using Dirichlet boundary conditions. We highlight that in the lattice formulation, if one does not correctly account for the energy contributions from the boundaries in the symmetrical tube geometry, then one would incorrectly find a repulsive Casimir potential as can be seen directly from the slope of the potential Fig.\ (\ref{fig:tube_total_S_Slope}).
    \item The functional form of the total Casimir potential (not per unit surface area) shown in Fig.\ (\ref{subfig:tube_ETotVac_Slope}) appears linear as compared to the form in Eq.\ (\ref{eqn:vcas_fit_plates}) for the case of parallel plates. Hence the Casimir force experienced by the surfaces of the tube diminishes slower with increasing separation distance.
    \item The potential of the tube at varying inverse coupling forms a smooth curve, but it does not collapse to a single curve as was the case for parallel plates. There are no additional energy contributions to the system that could introduce a different scaling, and we note this as a subject for further investigations. 
\end{itemize}

\subsubsection{Symmetrical Box}
The last geometry that we discuss for the non-abelian gauge theory in SU(3) is a hollow box shown in Fig.\ (\ref{subfig:box_geometry}). The box has finite extents on the three-dimensional spatial axis with side-lengths $R_x=R_y=R_z=R\sqrt{\sigma}$. We refer the reader to the preceding section for a discussion on the electromagnetic field-strength tensor components and the numerical field configurations results for the symmetrical box.

The geometry of a box requires the fixing of all spatial coordinates, leaving only the Euclidean time axis free. Consequently, all rotational symmetries are broken in this geometry because rotations in four-dimensional space require a minimum of two free axes. Therefore, as opposed to the geometries that we have discussed insofar, where we explored various symmetry relations to reduce the full expression of the energy density of the system into simpler expressions, such an approach is not possible here. 
%However, in figures (\ref{fig:fields3D_box_out} - \ref{fig:fields3D_box_face}), we showed that numerical results suggest that some field components are nevertheless equal. Hence it would be valuable in future work to explore whether this results in a general analytical simplification of the energy density expression.

The energy density of the box reads,
\begin{eqnarray}
    \varepsilon_{\text{box}}(x,y,z) &=& \frac{1}{2} \sum\limits_i \left[ \langle B_i^2 \rangle - \langle E_i^2 \rangle \right], \quad i=x,y,z
    \label{eqn:box_Edensity}
\end{eqnarray}
and is dependent on $x$, $y$ and $z$ because the gluon fields are enclosed in a finite volume (box) and the Casimir force acts on the walls of the box in all three spatial directions. The lattice expression of the energy density is a sum over the plaquettes on the worldvolume of the box,
\begin{eqnarray}
     \varepsilon(x,y,z)_{\text{box}}^{\text{lat}} &=& \langle S_{P_{ij}} \rangle = \frac{1}{N_{\tau}}\sum\limits_{N_{\tau}} S_{P_{ij}},
\end{eqnarray}
which comprise of all the plaquettes in the spatial and temporal directions according to Eq.\ (\ref{eqn:box_Edensity}). The resulting expression for the Casimir potential of the box is,
\begin{eqnarray}
    V^{\text{lat}}_{\text{Cas}}(R) &=& \left[ \int_{d\mathcal{V}}\, \varepsilon(x,y,z)_{\text{box}} \right]_{R-R_0} \\&=& \left[ \sum\limits_{N_{x}, N_{y} , N_{z}} \varepsilon(x,y,z)_{\text{box}}^{\text{lat}} \right]_{R-R_0}\\
    &=&  \sum\limits_{N_{x}, N_{y}, N_{z}} \left[ \langle S_{P_{ij}} \rangle_R - \langle S_{P_{ij}} \rangle_{R_0} \right],
    \label{eqn:box_casimir_lattice}
\end{eqnarray}
also represented by $\langle \langle S_{P_{ij}} \rangle \rangle$, and the expressions for the physical energies follow similarly to other cases already discussed in four-dimensional space.

\begin{figure}[!htb]
    %\begin{framed}
    \centering
    \subfigure[Vacuum Subtracted Fit]{{\includegraphics[scale=0.5]{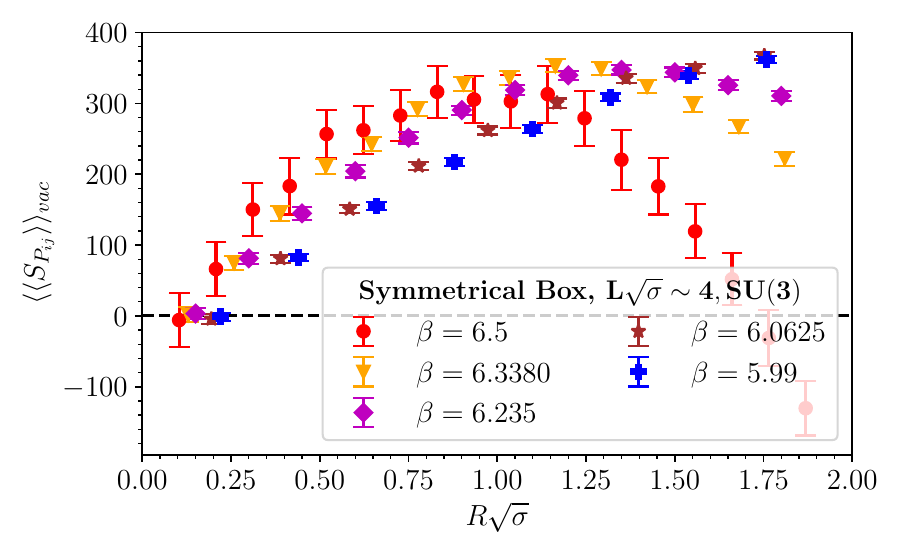} \label{fig:box_Svac} }}%
    \hspace{-0.45cm}
    \subfigure[$R_{\infty}$ Subtracted Fit]{{\includegraphics[scale=0.5]{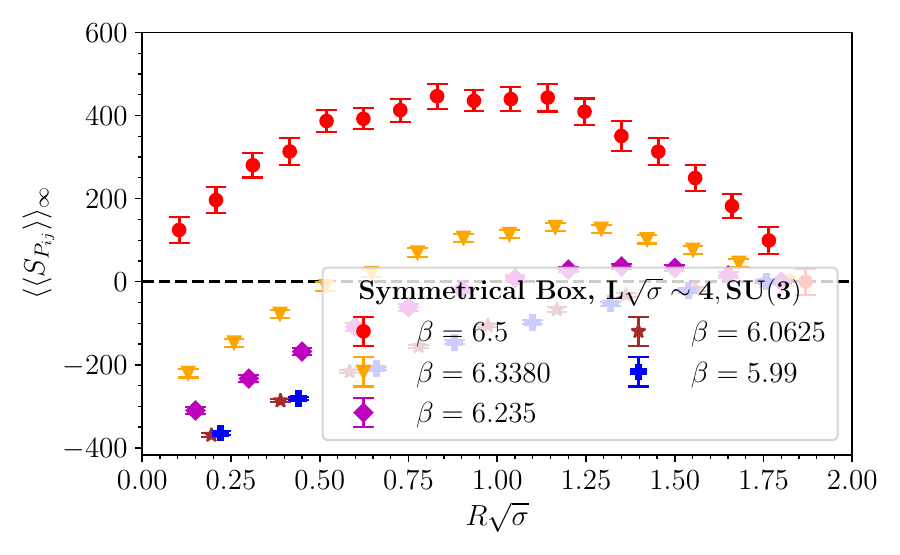} \label{fig:box_Sinf} }}%
    \caption{Total energy of the system in a symmetrical box in $(3+1)$D SU(3) for different couplings and different normalisation schemes in lattice units.}%
    \label{fig:box_total_S}
    %\end{framed}
\end{figure}

In the previous subsection, we discussed the intricacies of accounting for the energy contributions from the boundaries of the tube. Such difficulties were overcome by exploiting the linear dependence of the energy of the walls of the tube with respect to tube size. We encounter the same problem in the case of the box geometry, and once again, have to redefine the normalisation condition at $R_0$. 
%We now discuss these boundary energy contributions separately for the symmetrical and asymmetrical box.

We start by approaching this boundary problem for the symmetrical box in similar fashion as the symmetrical tube. We compute the lattice total energy of the system using Eq.\ (\ref{eqn:box_casimir_lattice}) and compare the result obtained using the vacuum subtraction and it's $R_{\infty}$ counterpart where we subtract the energy of a larger, symmetrical outer box of side-length, $L\sqrt{\sigma}/2$. These results are shown in Fig.\ (\ref{fig:box_total_S}), and we observe that the energy dependence does not vary linearly with increasing box size as was the case for the tube, allowing for the simplification of the calculation.
%from glancing at the system's energy dependence on the separation distance, it is already apparent that we will run into problems.\\

The main difference in the observed total energy in the symmetric box is that it is non-monotonous. We have already seen in the previous subsection that the energy contribution from the boundaries increases with the size of the boundary, i.e., the larger the box, the larger the energy contribution from the introduction of its walls should be. Hence this is a puzzling result because intuitively, the Casimir effect should vanish with increasing separation distance or box-size.

%While there is a possibility of the appearance of finite volume effects due to the confinement of modes inside the box, 
Nothing suggests that any other effects should materialise when the box is increased to a certain size. The same argument is true for thermal fluctuations, which should not fit inside the box. Therefore, we maintain that the observed effect of the total energy in the symmetric box is purely a consequence of the box's boundaries which include non-trivial contributions from the edges and corners. This argument is motivated by the $R_{\infty}$ subtracted energy in Fig.\ (\ref{fig:box_Sinf}), which vanishes at $R=L\sqrt{\sigma}/2$.

In the case of the symmetrical tube we observed that the functional form of the system's energy was linear for $R\to \infty$ for both normalisation schemes. However, in the case of the symmetrical box, the dependence of the total energy of the system on the separation distance (and consequently the expanding volume of the box) is more intricate. Perhaps a parabolic analytic form would work based on the observation of the energy at coupling, $\beta=6.5$, but we would need to increase our physical lattice volume to validate whether this can be generalised to other couplings. 

Despite the difficulty in assuming a simple analytical form for the system's energy, there are some common and consistent features. In the vacuum normalisation, in the limit, $R\sqrt{\sigma} \to 0$, the total energy of the system,
\begin{equation}
    E_{\text{Tot}}^{\text{Vac}} = E_{\text{Cas}} + E_{\text{Box}},
\end{equation}
approaches zero because in this limit, the energy contribution from the walls of the box vanish and so does the Casimir energy since it has to vanish in the absence of the box. On the other hand, in the $R_{\infty}$ subtraction scheme, in the limit $R\sqrt{\sigma} \to L\sqrt{\sigma}/2$, the total energy of the system given in Eq.\ (\ref{eqn:tube_Etot_infty}) with `Tube' replaced by `Box', vanishes. This is consistent with our expectations because in this limit, the Casimir energy should be negligible and if the `inner box' is the same size as the `outer box', then their energy contributions cancel. 
%At this juncture, in an attempt to find ways to describe the energy contributions from the expanding walls of the box, one can explore the rich ensemble of analytical and numerical functions out there. In any case, this does not appear to be a straightforward problem.\\

%Instead of trying to find relevant functional forms to describe the numerical data, 
We propose an alternative normalisation scheme for the symmetrical box. Each box is formed by a combination of six sides/faces, therefore, we start by finding the energy contribution of two finitely extending faces of the same size placed a distance, $R= L\sqrt{\sigma}/2$ apart. As discussed in previous sections, the energy contribution of the faces is independent of their separation distance. We merely place these faces far apart to ensure that there are no Casimir energy contributions to the total energy of the system.

\begin{figure}[!htb]
%\begin{framed}
\begin{center}
\includegraphics[scale=.5]{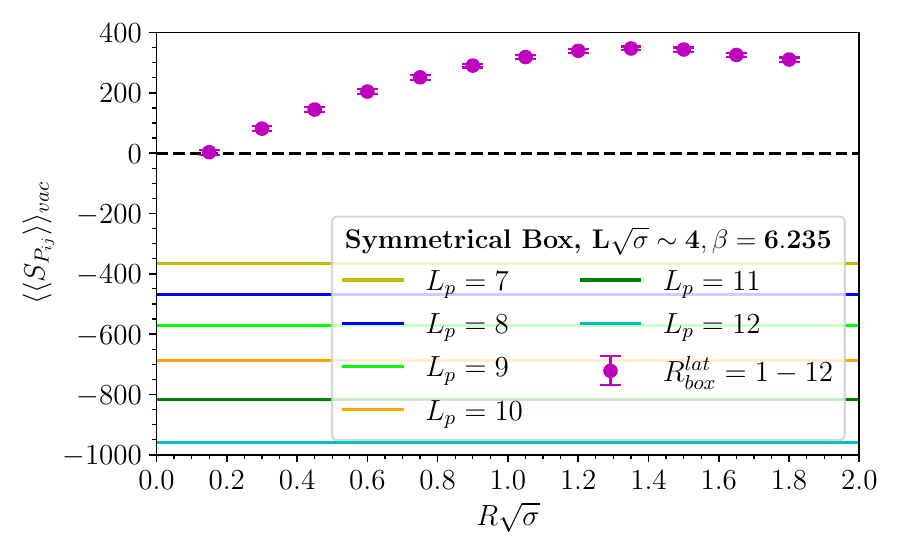}
\caption{Vacuum normalised energy of six faces compared to the vacuum normalised energy of a box with side-lengths equal to the size of the faces.}
\label{fig:Svac_box_plates}
\end{center}
%\end{framed}
\end{figure}

The box system has total energy,
\begin{eqnarray}
  E_{\text{Box}}^{\text{Tot}}  &=& E_{\text{Cas}}^{\text{Box}} + E_{\text{Box}} + E_{\text{Vac}},
  %\label{eqn:box_energy}
\end{eqnarray}
whereas the two finite parallel faces system has total energy,
\begin{eqnarray}
  E_{\text{Faces}}^{\text{Tot}}  &=& \cancelto{0}{E_{\text{Cas}}^{\text{Faces}}} + E_{\text{Faces}} + E_{\text{Vac}}.
  %\label{eqn:box_energy}
\end{eqnarray}
We numerically compute the energy contribution from two finite faces placed far apart. In the end we are interested in the six-faces configuration in order to compare the total energy of this configuration to the energy of the box system (whose geometry is formed by a total of six faces). Thus, we multiply $E_{\text{Faces}}^{\text{Tot}}$ by three to obtain,
\begin{eqnarray}
  E_{\text{Faces}}^{\text{Tot}_2}  &=& 3E_{\text{Faces}}^{\text{Tot}} = 3(E_{\text{Faces}} + E_{\text{Vac}}) \\&=& \Tilde{E}_{\text{Faces}} + 3E_{\text{Vac}}.
  %\label{eqn:box_energy}
\end{eqnarray}
Performing a vacuum subtraction on the energies of the two systems we obtain,
\begin{eqnarray}
  \Tilde{E}_{\text{Box}}^{\text{Tot}}  &=& E_{\text{Box}}^{\text{Tot}} - E_{\text{Vac}} = E_{\text{Cas}}^{\text{Box}} + E_{\text{Box}} \label{eqn:box_energy_vacnorm}\\
  \Tilde{E}_{\text{Faces}}^{\text{Tot}}  &=& E_{\text{Faces}}^{\text{Tot}_2} - 3E_{\text{Vac}} = \Tilde{E}_{\text{Faces}},
  %\label{eqn:box_energy}
\end{eqnarray}
and we show the vacuum normalised energy for the box in Fig.\ (\ref{fig:Svac_box_plates}) as coloured points, along with the corresponding energy of the six faces forming a box as coloured lines, where $R=L_p$ is the lattice side-length of each face. It is reasonable that the energy contributions from the faces increases with the size of the faces.

%\noindent
%Note that the energy contribution from six faces is not equal to the energy contribution of the box (composed of six faces). Specifically, the energy from creating six faces is less than the energy from creating a box with the same side-lengths. Hence, this is not a one-to-one comparison of the energy contributions because the geometrical set-up of small parallel faces is different from that of a single box with the same side-lengths. The differences in energies can be attributed to the presence of corners in the box (see the field contributions from the corners in the preceding chapter) and a non-trivial contribution from the edges in each set-up.\\

Note that the energy contribution from six faces is not equal to the energy contribution of the box (composed of six faces). Hence, this is not a direct comparison of the energy contributions in the two systems because the geometrical set-up of finite parallel faces is different from that of a single box with the same side-lengths. The differences in energies can be attributed to the presence of corners in the box (see the field contributions from the corners in the preceding section) and varying contributions from the edges in the two geometries.

\begin{figure}[!htb]
%\begin{framed}
\begin{center}
\includegraphics[scale=.5]{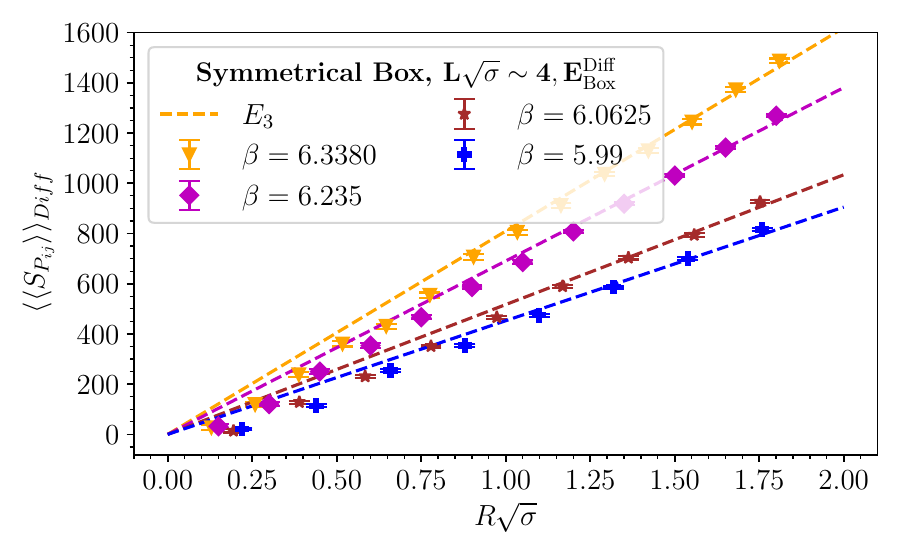}
\caption{Energy difference between the creation of six faces and a box with side-lengths equal to the size of the faces.}
\label{fig:Sdiff_box_plates}
\end{center}
%\end{framed}
\end{figure}

The energy difference between the two systems (i.e.\ a box compared to a combination of six faces) is given by, 
\begin{equation}
    E_{\text{Box}}^{\text{Diff}}(R) = \Tilde{E}_{\text{Box}}^{\text{Tot}} - \Tilde{E}_{\text{Faces}}^{\text{Tot}} = E_{\text{Cas}}^{\text{Box}} +  E_{\text{Box}} - \Tilde{E}_{\text{Faces}},
\end{equation}
and shown in Fig.\ (\ref{fig:Sdiff_box_plates}) for different gauge couplings. The Casimir energy contribution for the geometry with two finite faces is negligible irrespective of the size of the faces because the faces are placed far apart at a distance, $R= L\sqrt{\sigma}/2$. At large separation distances (large enough box-size), the Casimir energy for the box becomes negligible, leaving only the energy difference between the boundaries of the two geometries. Therefore, for $R\gtrsim 1$, this energy difference should describe,
\begin{equation}
    E_3 = \left . E_{\text{Box}}^{\text{Diff}}\right|_{R\gtrsim 1} = E_{\text{Box}} - \Tilde{E}_{\text{Faces}}.
\end{equation}

We find that this energy difference varies linearly with the size of the box and faces. Therefore, instead of finding intricate functions to extract the energy of the box's boundaries from the observed behaviour in Fig.\ (\ref{fig:box_total_S}), we have reduced the problem to finding the fitting parameters of a linear function. We find the constant of the fit by fixing the energy difference at the $R=0$ endpoint to zero since the boundary contributions from both geometries vanish. The resulting fit parameters are provided in Table (\ref{tab:Sdiff_fit_params}) at different couplings.

\begin{table}[!htbp]
\begin{ruledtabular}
    \centering
        %{\rowcolors{2}{green!80!yellow!50}{green!70!yellow!40}
        \begin{tabular}{ cccc}
        %\hline
        %\multicolumn{3}{|c|}{Country List} \\
        %\hline
        Energy & $\mathbf{\beta}$ & $m$ & $c$\\
        \hline
         & 6.3380 & 809.36 & 0\\
        $\langle \langle S_{P_{ij}} \rangle \rangle_{Diff}$ & 6.235 & 690.41 & 0 \\
         & 6.0625 & 516.86 & 0 \\
         & 5.99 & 452.55 & 0\\
        %\hline
        \end{tabular}%}
    \caption{Linear fit parameters for the energy difference between the boundaries of six faces and those of a symmetrical box with equal side-lengths as the faces.}
    \label{tab:Sdiff_fit_params}
\end{ruledtabular}
\end{table}

Once we have obtained this energy difference, we can normalise the energy of a symmetrical box using the energy from the boundaries of a finite parallel faces configuration, accounting for six faces that form a box. We then use the linear fit to remove the expected energy difference between the boundaries of a box and the parallel faces set-up,
\begin{eqnarray}
  E_{\text{Cas}} &=& \Tilde{E}_1 - \Tilde{E}_2 - E_3.
  \label{eqn:box_cas_energy}
\end{eqnarray}
This approach allows us to isolate the Casimir energy of the symmetrical box, as well as the approximate energy from the walls of the box,
\begin{eqnarray}
  E_{\text{Box}} &=& \Tilde{E}_2 + E_3,
  \label{eqn:box_walls_energy}
\end{eqnarray}
where $E_1$ is the total energy of the system with a box present, $E_{\text{Faces}}^{\text{Tot}}$ is the total energy of the system with two parallel faces placed far apart (times three) and $E_3$ is the energy difference of the respective boundaries.

\begin{figure}[!htb]
%\begin{framed}
\begin{center}
\includegraphics[scale=.5]{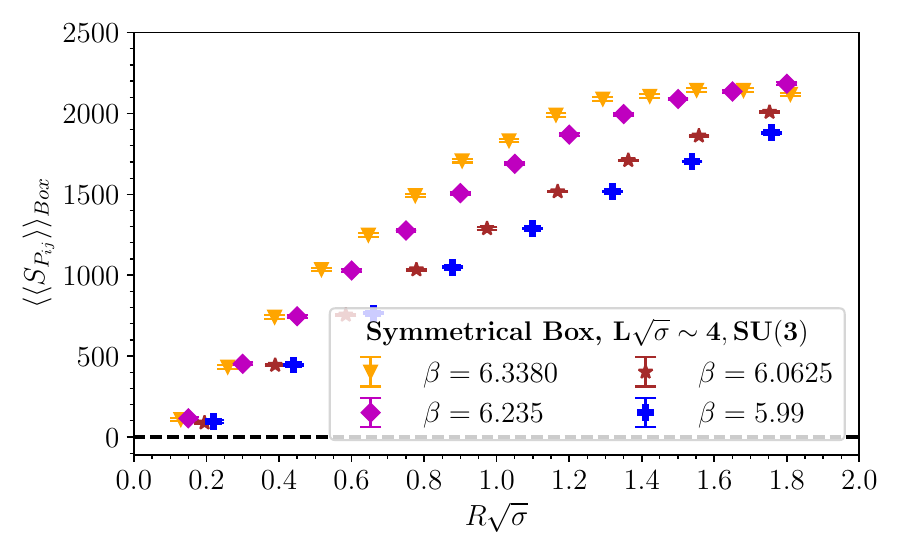}
\caption{The total energy contribution from the boundaries of a symmetrical box of different sizes.}
\label{fig:Sbox_creation}
\end{center}
%\end{framed}
\end{figure}

Most importantly, we have not only isolated the Casimir energy for the box, but also, the energy contributions from the walls of the box given by Eq.\ (\ref{eqn:box_walls_energy}). Despite not having an analytical formula to describe the energy from creating the box, we provide an approach to numerically extract it from lattice simulations data and we show this energy in Fig.\ (\ref{fig:Sbox_creation}). Although our result is sensitive to the fit parameters, it provides a reasonable estimate of the Casimir effect in this geometry.

\begin{figure}[!htbp]
    \centering
    \subfigure[Total Casimir Potential\label{subfig:VcasTotal_box}]{{\includegraphics[width=0.9\linewidth]{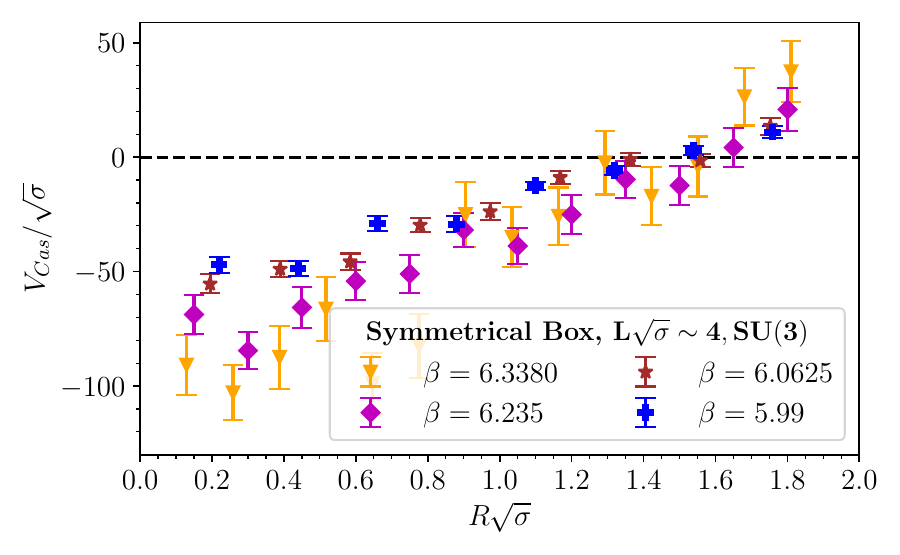} }}%
    \hspace{1em}
    \subfigure[Potential/Surface Area\label{subfig:Vcas_box_area}]{{\includegraphics[width=0.9\linewidth]{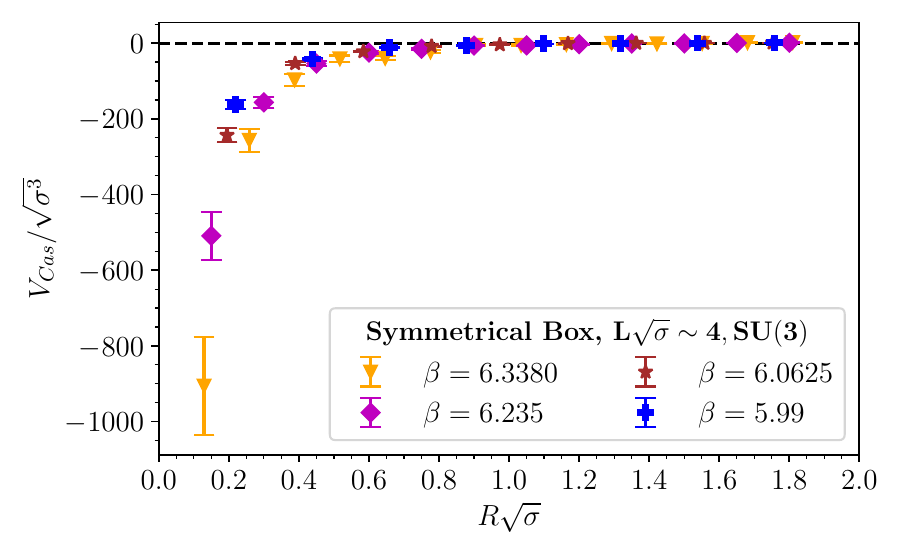} }}
    %\hspace{1em}
    \caption{The total Casimir potential for a symmetrical box with side lengths $R\sqrt{\sigma}$ in SU(3).}
    \label{fig:Vcas_box}
\end{figure}

The resulting absolute Casimir potential and the potential per unit surface area of a symmetrical box is shown in Fig.\ (\ref{fig:Vcas_box}), where the area is given by, $A=6R^2$. One key feature of this potential is that the Casimir force experienced by the symmetrical box is \textit{attractive}. This result is consistent with the Casimir effect of a weakly coupled, massless non-interacting scalar field computed in Ref.\ \cite{Mogliacci:2018oea}. As shown in Fig.\ (\ref{fig:box_Sinf}), incorrectly accounting for the boundary energy contributions for the box would result in a repulsive Casimir effect for some couplings and an attractive potential for others, with a sign flip in-between.\\

%%%%%%%%%%%%%%%%%%%%%%%%%%%%%%%%%%%%%%%%%%%%%%%%%%%%%%%%%%%%%%%%%%%%%%%%%%%%%%%%%%%%%%%%%%%%%%%%%%%%%%%%%%%%%%%%%%%%%%

%%%%%%%%%%%%%%%%%%%%%%%%%%%%%%%%%%%%%%%%%%%%%%%%%%%%%%%%%%%%%%%%%%%%%%%%%%%%%%%%%%%%%%%%%%%%%%%%%%%%%%%%%%%%%%%%%%%%%%%%

%%%%%%%%%%%%%%%%%%%%%%%%%%%%%%%%%%%%%%%%%%%%%%%%%%%%%%%%%%%%%%%%%%%%%%%%%%%%%%%%%%%%%%%%%%%%%%%%%%%%%%%%%%%%%%%%%%%%%%%

\subsection{The Polyakov Loop and Deconfinement}
The goal of this section is to understand the temperature dependence of the Casimir potential. It is suggested in Ref.\ \cite{Chernodub:2018pmt} that in the two-dimensional non-abelian gauge theory, where the Casimir effect was studied for parallel wires in SU(2), the region between the wires is a Casimir-induced deconfined phase. As a result, the gluons in this finite region exhibit similar behaviour to thermal glueballs in a heat-bath at finite temperature. This is attributed to the absence of the periodicity of gluonic fields in the presence of Casimir boundaries. The same behaviour is observed for parallel plates in SU(3) \cite{Chernodub:2023dok} where these thermal glueballs are interpreted as colourless states of gluons bound to their negatively coloured images on the chromoelectric boundary.

This suggestion is also supported by evidence of the measured mass of the relevant degrees of freedom in the Casimir interaction, $M_{\text{Cas}}$ in Eq.\ (\ref{eqn:vcas_fit}). The Casimir mass was shown to assume a lower value than the lightest ground state glueball in SU(2) \cite{Chernodub:2018pmt} and SU(3) \cite{Chernodub:2023dok}, which should be the lowest mass in the pure gluonic system. The reduced glueball mass in the `heat-bath' Casimir cavity is inconsistent with finite temperature lattice measurements of thermal glueball properties, which show that the pole mass of the glueball ground state does not change below $T_c$, but increases with temperature above $T_c$ \cite{Arikawa:2025kjx}. This glueball mass dependence above $T_c$ is opposite to the results of previous studies \cite{Ishii:2002ww,Meng:2009hq}.
%The reduced glueball mass in the `heat-bath' Casimir cavity is consistent with finite temperature lattice measurements of thermal glueball properties, which show that the mass of the lightest glueball, $M_{0^{++}}$, decreases with temperature, even in the confined phase, with a mass reduction of up to $\sim 20\%$ around $T_c$ in SU(3) \cite{Ishii:2002ww,Meng:2009hq}.

The discussion of the Casimir mass in the non-abelian theory for parallel wires in (2+1)D and plates in (3+1)D for both SU(2) and SU(3) is expanded on in Ref.\ \cite{Ngwenya:2025mpo}. Similarly, a Casimir mass lower than the lightest glueball, $M_{0^{++}}$, in the corresponding theories at temperatures corresponding to the confined phase is obtained. While we have studied the Casimir effect for the tube and box geometries, at the present stage, we have not performed Casimir mass calculations to test for any geometry dependence in the three-dimensional theory. %This suggests that the boundary induced deconfined phase exists in all geometries. While we have studied the Casimir effect for the tube and box geometry, at the present stage, we have not performed Casimir mass studies to test for any geometry dependence. \\

\begin{figure}[!htbp]
    \centering
    \subfigure[\label{subfig:Vcas_temp}]{{\includegraphics[width=0.9\linewidth]{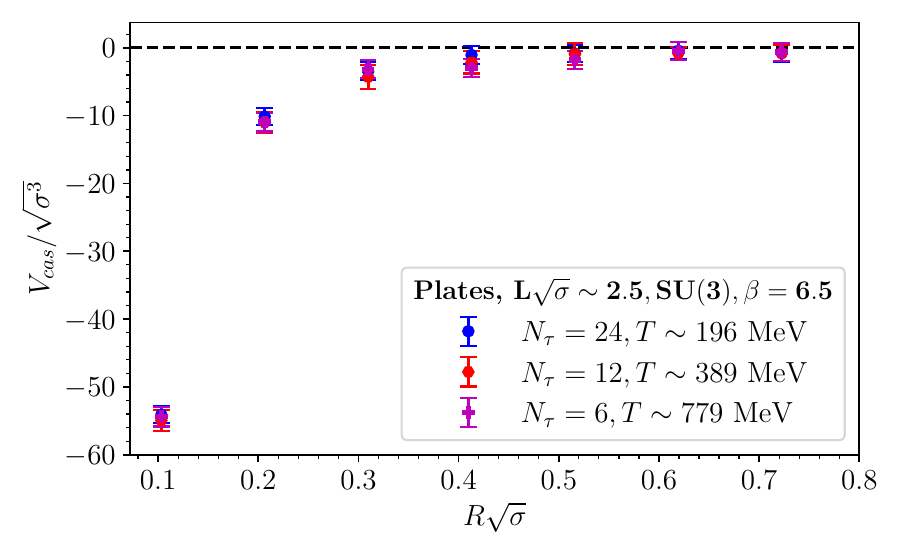} }}%
    \hspace{1em}
    \subfigure[\label{subfig:Vcas_temp_ratio}]{{\includegraphics[width=0.9\linewidth]{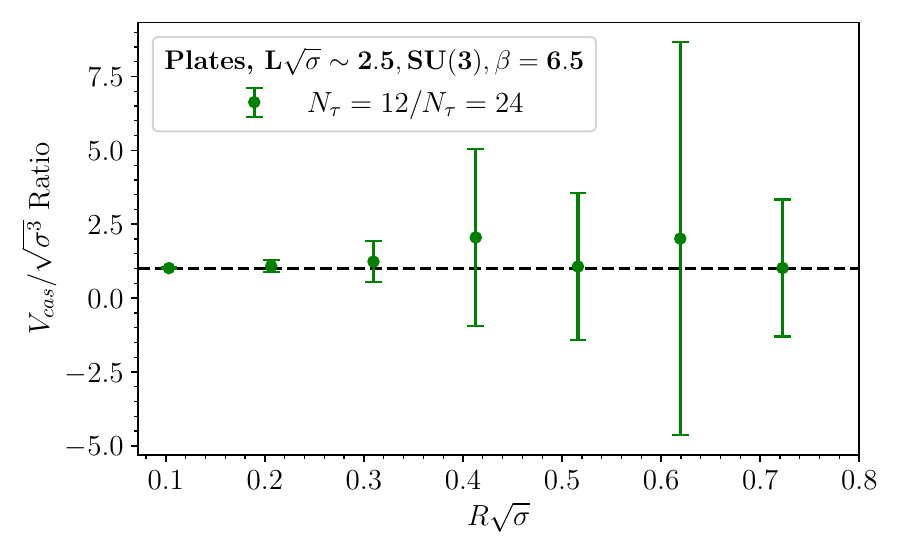} }}
    %\hspace{1em}
    \caption{\label{fig:Vcas_plates_temp}The Casimir potential per unit area between two parallel plates separated by a distance $R\sqrt{\sigma}$ in SU(3) at different temperatures.}%
\end{figure}

In Fig.\ (\ref{fig:Vcas_plates_temp}), we show the temperature dependence of the measured Casimir effect of the parallel plate geometry as we move from the number of temporal lattice grid points $N_{\tau}$ associated with the confined to deconfined phase. Because the region between the plates is already a boundary induced deconfined phase where the relevant degrees of freedom have different properties compared to those measured for pure gluodynamics, decreasing $N_{\tau}$ (increasing the temperature) from the confined to deconfined phase does not alter the Casimir effect. The non-abelian Casimir effect is a deconfined phase phenomenon due to the boundaries and is insensitive to temperature changes across $T_c$. However, it may be valuable in future studies to explore the limit $T\to 0$, which is resource intensive on the lattice.

In order to better understand gluodynamics along the temperature axis, we look at the finite temperature deconfinement order parameter described by the Polyakov loop (or \textit{thermal Wilson line}) \cite{Polyakov:1976fu}, 
\begin{equation}
    L_{\bm x} = \frac{1}{N_c} \text{Tr} \prod_{\tau=0}^{N_{\tau -1}} U_{\mu=4}(\bm x,\tau),
\end{equation}
which is a straight-line product of $N_{\tau}$ link variables in the Euclidean time direction, all situated at a single spatial point $\bm x \coloneqq (x_1, x_2, x_3)$. The Polyakov loop is a trace over a closed loop (through periodic boundary conditions) and is therefore gauge invariant. 
%In lattice terminology, the term under the trace is also referred to as the \textit{temporal transporter} and appears in the definition of the Wilson loop.\\

The deconfinement order parameter is given by the spatial average
\begin{equation}
    %L = \frac{1}{V_r} \left\langle \left|\sum_{\bm x \in V_r} L_{\bm x} \right| \right\rangle,
    L \equiv \frac{1}{V_r} \left\langle \sum_{\bm x \in V_r} L_{\bm x}  \right\rangle,
    \label{eqn:polyakov_loop_avg}
\end{equation}
where $V_r=N_s^3$ is the physical volume. Given that we impose boundaries, note that we use the subscript $r$ to specify between the regions in our Casimir effect geometries, where one region is inside (e.g., between the plates) and the other is outside. Thus we compute the Polyakov loop separately between these two regions.

The expectation value of a single Polyakov loop $L_{\bm x}$ in infinite volume is interpreted as the probability to observe a single static charge and is approximated by the exponential of the \textit{free energy}, $F_q$ \cite{Gattringer:2010zz},
\begin{equation}
    \frac{1}{V}\left| \left\langle \int d^3\, x L_{\bm x} \right\rangle \right| \sim e^{-F_q/T},
\end{equation}
of a single static charge. 
%That is, the theory is confined in the limit $F_q \to \infty$, implying that isolated quarks cannot exist as free particles in this phase, and deconfined for finite free energy. Therefore, $\langle L \rangle = 0$ in the confined phase and $\langle L \rangle \neq 0$ in the deconfined phase.

\begin{figure}[!htb]
    \centering
    \subfigure[$T\simeq 195$ MeV\label{subfig:poly_Nt24}]{{\includegraphics[width=0.9\linewidth]{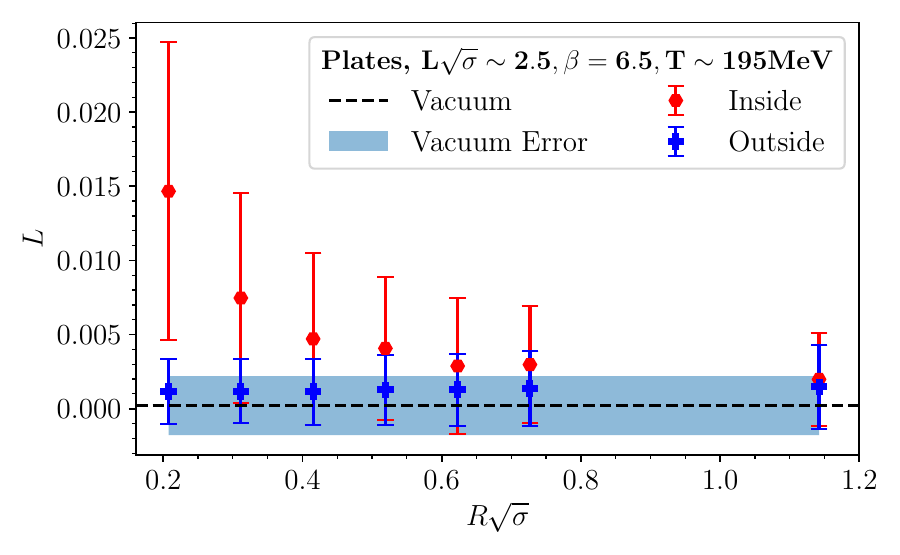} }}%
    \hspace{1em}
    \subfigure[$T\simeq 389$ MeV\label{subfig:poly_Nt12}]{{\includegraphics[width=0.9\linewidth]{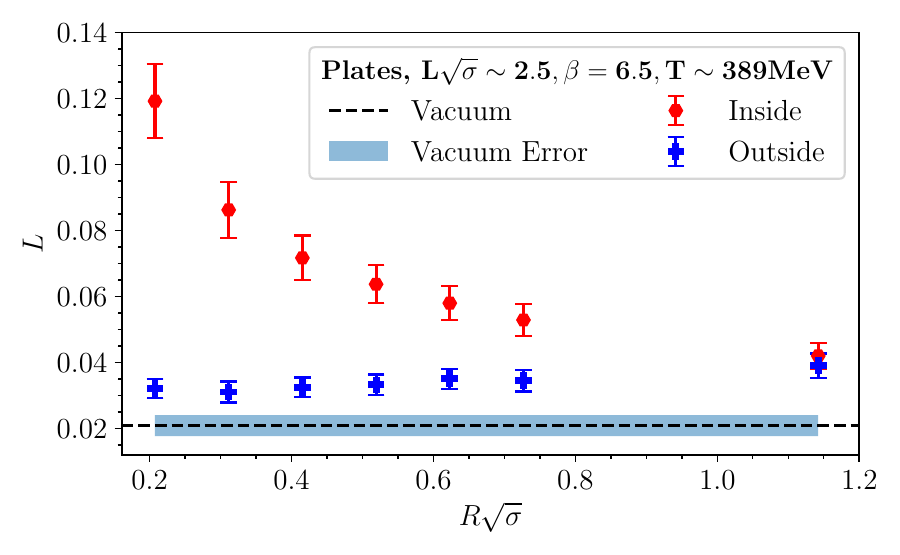} }}
    \hspace{1em}
    \subfigure[$T\simeq 779$ MeV\label{subfig:poly_Nt6}]{{\includegraphics[width=0.9\linewidth]{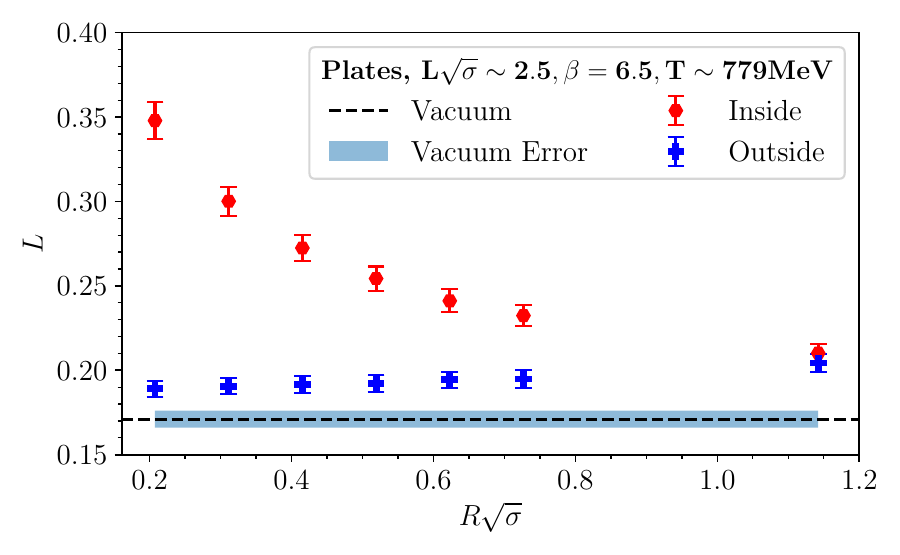} }}
    %\hspace{1em}
    \caption{The Polyakov loop expectation value in the region between and outside the parallel plates in SU(3) in the confined and deconfined phase.}%
    \label{fig:poly_deconfined}
\end{figure}

We have shown in Fig.\ (\ref{fig:Vcas_plates_temp}) that increasing the temperature from the confined to deconfined phase has no observable effect on the Casimir effect between parallel plates in SU(3). We now look at the corresponding Polyakov loop expectation values as $N_{\tau}$ is decreased from the confined to deconfined phase. In Fig.\ (\ref{subfig:poly_Nt24}), we show the Polyakov loop expectation value in the confined phase as defined in Eq.\ (\ref{eqn:polyakov_loop_avg}). In the absence of the plates, $L\approx 0$ because the pure glueball system is in the confined phase and center symmetry is preserved.

In the region between the plates, the Polyakov loop expectation value is non-zero and increases with decreased separation distance between the plates, supporting the idea of a deconfined phase in this region. While there is a clear trend in the Polykov loop expectation value, the error bars are large resulting in data-points for $R>1$ being consistent with zero, which would suggest the system remains confined at these distances. Improvement of the statistics is necessary to draw quantitative conclusions. Meanwhile the Polyakov loop, $L\sim 0$ in the large volume outside the plates where the gluonic system remains confined and increases only slightly with separation distance. As the separation distance between the plates, $R\to L\sqrt{\sigma}/2$ where the boundaries are mirrored, the expectation value of the Polykov loop inside and outside the plates approaches the same value because the two regions become indistinguishable.

In Fig.\ (\ref{fig:poly_deconfined}), as the temperature is increased and the system moves into a deconfined phase, the vacuum (in the absence of the plates) Polyakov loop expectation value in the pure gluonic system becomes non-zero, thus signalling deconfinement. While in the deconfined phase, this vacuum expectation value continues to increase with temperature, however $L_{vac} < L_{in}$. In the confined phase, we observed that in the region outside the plates, $L\approx 0$ and equal to the vacuum expectation value. This is no longer the case in the deconfined phase, $L_{vac} < L_{out} < L_{in}$ and there is an observable linear increase in $L_{out}$ with separation distance. 

On the other hand, in the region between the plates, the magnitude of the deconfinement order parameter increases with temperature. However we have observed that this does not effect the Casimir effect. Given the observable change in the Polyakov loop with increasing temperature, it is surprising that the Casimir effect remains insensitive to the temperature change. Thermal glueballs studies \cite{Meng:2009hq} show that the glueball mass decreases with increasing temperature above $T_c$, hence the relevant degrees of freedom in the Casimir interaction should change with increased temperature. This temperature independence of the Casimir effect requires further investigation.

\begin{figure}[!htb]
\centering
\includegraphics[width=0.9\linewidth]{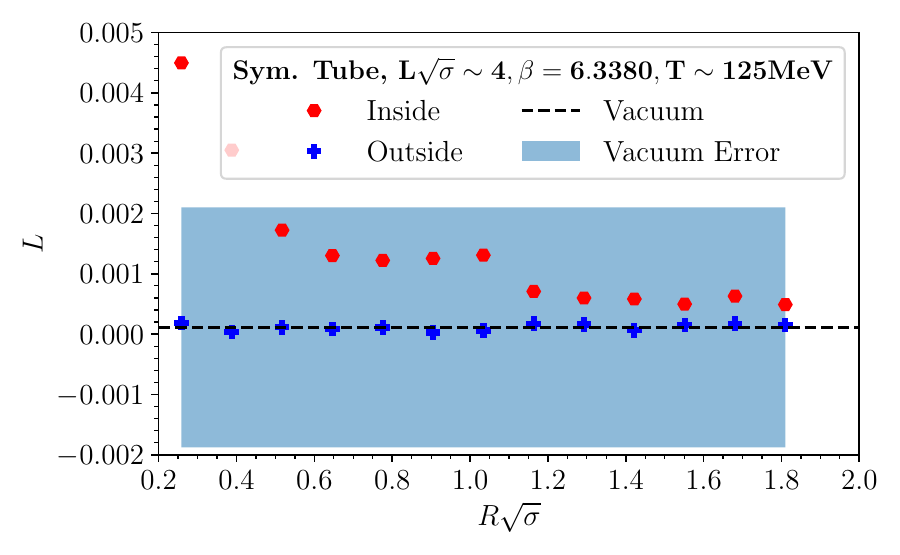}
\caption{The Polyakov loop expectation value in the region inside and outside a symmetrical tube in SU(3) in the confined phase.}%
\label{fig:poly_tube_confined}
\end{figure}

Lastly, we show the expectation value of the Polyakov loop in the confined phase for the symmetrical tube in Fig.\ (\ref{fig:poly_tube_confined}), omitting the large error-bars for neatness. We emphasize that these results are not statistically significant but hint at an interesting trend. The Polyakov loop for both the symmetrical and asymmetrical tube is discussed in Ref.\ \cite{Ngwenya:2025mpo}. In both geometries, the Polyakov loop fluctuates around the vacuum expectation value outside the tube. Inside the symmetrical tube, the Polyakov loop behaves similarly to the parallel plate geometry and approaches zero with increased tube-size.

%\begin{figure}[!htbp]
%\centering
%\includegraphics[width=0.9\linewidth]{Charm_v2GbDp3040_interaction_comparison.png}
%\caption{\label{fig:charmv2IntOff}Charm quark $v_{2}(p_T)$ at $\sqrt{s_{NN}} =5.5$ TeV for the 30-40\% centrality class with the interaction between energy loss and flow on (blue) and off (red) for the EE D(p) scenario.}
%\end{figure}

\section{\label{Conclusions}Conclusions and Outlook}
The Casimir effect has been successfully studied primarily in abelian gauge theory using perturbative analytic methods, complex geometries require simplifications such as fixed boundary conditions such as Dirichlet, Neumann etc. These simplifications in turn lead to inaccuracies in the predicted Casimir effect, for example thermal corrections \cite{Mitter:1999hu, Decca:2003td, Ghisoiu:2010fga}. These difficulties encountered in perturbative techniques provide a strong motivation for the exploration of the Casimir effect using non-perturbative methods such as lattice QED/QCD. 

We have presented Casimir potential results in (3+1)D non-abelian gauge theories for the geometries of parallel plates in SU(3) shown in Fig.\ (\ref{fig:casimir_plates}), as well as SU(3) results for a symmetrical tube in Fig.\ (\ref{fig:Vcas_tube}) and a symmetrical box in Fig.\ (\ref{fig:Vcas_box}). We show that the resulting Casimir potential is \textit{attractive} in all these geometries. 
%Our work follows a study of the Casimir effect in non-abelian gauge theory on the lattice for perfectly conducting static parallel wires in (2+1)D SU(2) at zero temperature using chromoelectric boundaries is presented in Ref.\ \cite{Chernodub:2018pmt}. 

In order to better understand the implication of these boundary conditions on the gauge fields, we have studied the effect of the chromoelectric boundary conditions on the individual field components in section (\ref{chapter:boundaries_em_fields}). We find different degrees of change in the magnitude of the individual field components at the boundaries and the surrounding region. 
%Analytically applying rotational matrices on the field-strength tensor, we find that there exists an equivalence in some of the electric and magnetic field components based on the geometrical set-up due to the rotational symmetries. Using these rotational symmetries, we numerically show the equivalent field expectation values for the configurations of the parallel plates and the symmetrical tube. This, in turn allows us to simplify the expressions of the energy density in these geometries due to cancellations in the equivalent terms. Meanwhile, there are no rotational symmetries to explore in the box geometry.

In the case of parallel plates, increasing the number of degrees of freedom from $N_c:2\to3$ results in an increased Casimir potential for intermediate separation distances. However, we observe that at very small and large separation distances, the potential is greater in SU(2). At this point, we have no physical interpretation of this phenomena, and the observed behaviour could instead point out the need to better understand the analytical Casimir fitting functions. 

%Overall, the measured Casimir mass in both the two and three-dimensional theory is lower than the mass of the lightest ground state glueball, $M_{0^{++}}$ which is the lowest mass in a pure gauge non-abelian system. The low Casimir mass indicates that the relevant degrees of freedom in the Casimir interaction are not glueballs, but rather lighter particles exhibiting similar behaviour to thermal glueballs in a heat-bath.\\

Due to the changing area of the plates with separation distance in the tube and box geometry as opposed to the parallel plate geometry where the area was fixed, we show that a careful treatment of the energy contribution from the boundaries is necessary, else one could incorrectly measure a repulsive potential for the symmetrical tube. Whereas, the potential for the symmetrical box would appear to move from attractive to repulsive depending on the lattice coupling and box-size. We have proposed two methods in which this energy of creating the boundaries could be accurately accounted in the geometry of a symmetrical tube employing linear fitting functions. Whereas the energy contribution from the boundaries of a symmetrical box is nonlinear and unique in that it is non-monotonous. The normalisation is performed by subtracting the energy of creating the six faces that make up the box.

We then show that the Casimir potential is independent of a temperature increase from the confined to the deconfined phase. We provide results of the expectation value of the Polyakov loop for the plates and symmetrical tube. These results confirm that at temperatures below $T_c$, the Polyakov loop is zero outside the plates (consistent with a confined system) but it is non-zero in the region between the plates. This observation confirms that the region between each geometry is a boundary-induced deconfined phase \cite{Chernodub:2018pmt}, hence the measured Casimir mass of the relevant degrees of freedom is lower than the lightest glueball masses.

Our study has only been performed at microscopic separation distances between the boundaries, which allows for the interpretation of the matter inside the boundaries as a boundary induced deconfined phase. However, in the case of macroscopic separation distances between the boundaries, the matter in the region between the boundaries has no knowledge of the presence of the boundaries, as is the case for the region outside the boundaries even for microscopic distances considered in this study. Hence at low temperatures, a plausible scenario would be the Polyakov indicating confined matter in both the region between and outside the boundaries, while showing signs of deconfinement close to the boundaries (i.e., a deconfined `boundary layer'). This suggests a possible alternative to the interpretation of the region inside the boundaries and we highlight it as an idea for future investigations. 

The data containing the findings of this study are available in the Zenodo repository \cite{ngwenya_2025_17188030}.

\begin{acknowledgments}
The authors wish to acknowledge the SA-CERN Collaboration and the South African National Research Foundation (NRF) for their generous financial contributions towards this work. This work also received support from ERASMUS+ project 2023-1-NO01-KA171-HED-000132068. We would also like to acknowledge the South African Centre for High Performance Computing (CHPC) and the University of Stavanger's computing cluster for their computational resources.
\end{acknowledgments}

\FloatBarrier %Block figures from going into the appendix

\twocolumngrid

% \bibliography{apssamp}% Produces the bibliography via BibTeX.
%\printbibliography
\bibliography{NgwenyaRothkopfHorowitz}
\bibliographystyle{apsrev4-2}

\end{document}